\documentclass[sn-mathphys-num]{sn-jnl}% Math and Physical Sciences Numbered Reference H$_2^{18}$O 
\usepackage{graphicx}%
\usepackage{multirow}%
\usepackage{amsmath,amssymb,amsfonts}%
\usepackage{amsthm}%
\usepackage{mathrsfs}%
\usepackage[title]{appendix}%
\usepackage{xcolor}%
\usepackage{textcomp}%
\usepackage{manyfoot}%
\usepackage{booktabs}%
\usepackage{algorithm}%
\usepackage{algorithmicx}%
\usepackage{algpseudocode}%
\usepackage{listings}%
\usepackage{multibib}
\newcites{main}{ References}
\newcites{supp}{Methods References}

\usepackage[version=4]{mhchem}
\usepackage{chemformula}

\usepackage[normalem]{ulem}

\usepackage{placeins}

\newcommand*\farcs{\ensuremath{\overset{\prime\prime}{.}}}

\usepackage{gensymb}

\usepackage{lmodern}
\usepackage{anyfontsize}

\begin{document}
	
	\title[Article Title]{Pristine ices in a planet-forming disk revealed by heavy water}% \newline Pristine water in planet-forming disk revealed through heavy isotope enrichment}
	
	%%=============================================================%%
	%% GivenName	-> \fnm{Joergen W.}
	%% Particle	-> \spfx{van der} -> surname prefix
	%% FamilyName	-> \sur{Ploeg}
	%% Suffix	-> \sfx{IV}
	%% \author*[1,2]{\fnm{Joergen W.} \spfx{van der} \sur{Ploeg} 
		%%  \sfx{IV}}\email{iauthor@gmail.com}
	%%=============================================================%%
	
	\author*[1]{\fnm{Margot} \sur{Leemker}}\email{margot.leemker@unimi.it}
	\author[2]{\fnm{John J.} \sur{Tobin}}%\email{iiauthor@gmail.com}
	\author[1]{\fnm{Stefano} \sur{Facchini}}
	\author[3]{\fnm{Pietro} \sur{Curone}}
	\author[4,5]{\fnm{Alice S.} \sur{Booth}}	
	\author[6,7]{\fnm{Kenji} \sur{Furuya}}
	\author[8]{\fnm{Merel L. R.} \sur{van 't Hoff}}	
	
	\affil*[1]{\orgdiv{Dipartimento di Fisica}, \orgname{Universit\`a degli Studi di Milano}, \orgaddress{\street{Via Celoria 16}, \city{Milano}, \postcode{20133}, \country{Italy}}}
	\affil[2]{\orgname{National Radio Astronomy Observatory}, \orgaddress{\city{Charlottesville}, \state{VA}, \country{USA}}}
	\affil[3]{\orgdiv{Departamento de Astronom\'ia}, \orgname{Universidad de Chile}, \orgaddress{\street{Camino El Observatorio 1515}, \city{Las Condes, Santiago}, \country{Chile}}}
	\affil[4]{\orgdiv{Center for Astrophysics}, \orgname{Harvard \& Smithsonian}, \orgaddress{\street{60 Garden St.}, \city{Cambridge}, \postcode{02138}, \state{MA}, \country{USA}}}	
	% \affil[5]{\orgname{Clay Postdoctoral Fellow}}
	\affil[6]{\orgdiv{Department of Astronomy}, \orgname{Graduate School of Science, University of Tokyo}, \orgaddress{\city{Tokyo}, \postcode{113-0033}, \country{Japan}}}
    \affil[7]{\orgname{RIKEN Pioneering Research Institute}, \orgaddress{\street{2-1 Hirosawa, Wako-shi}, \city{Saitama}, \postcode{351-0198}, \country{Japan}}}
	\affil[8]{\orgdiv{Department of Physics and Astronomy}, \orgname{Purdue University}, \orgaddress{\street{525 Northwestern Avenue}, \city{West Lafayette}, \postcode{47907}, \state{IN}, \country{USA}}}
	
	% List of co-authors in arbirtrary order: Stefano, Ewine, John, Pietro Curone, Alice, Merel, Kenji
	
	%%==================================%%
	%% Sample for unstructured abstract %%
	%%==================================%%
	
	\abstract{
        Water is essential to our understanding of the planet-formation process and habitability on Earth. Although trace amounts of water are seen across all phases of star and planet formation, the bulk of the water reservoir often goes undetected, hiding crucial parts of its journey from giant molecular clouds to planets. This raises the question of whether water molecules in comets and (exo-)planets is largely inherited from the interstellar medium or if the water molecules are destroyed and then reformed in the disk. Water isotopologue ratios involving doubly deuterated water (\ce{D2O}) are a sensitive tracer to answer this question. We present strong evidence of inheritance through an enhancement of \ce{D2O} in the outbursting V883 Ori disk. The high \ce{D2O}/\ce{H2O} ratio of $(3.2 \pm 1.2) \times 10^{-5}$ is consistent with values seen in protostellar envelopes and a comet and is two orders of magnitude higher than expected if water is reprocessed. The high deuteration of the heaviest isotopologues \ce{D2O}/HDO = $(2.3 \pm 1.0) \times $HDO/\ce{H2O} further establishes the inheritance of water. We conclude that water ice in disks originates from the earliest phases of star formation, providing the missing link between cold dark clouds and (exo-)comets.	
	}


	\maketitle

Water may have been delivered to Earth via cometary and/ or asteroid impacts, tracing the pristine material left over from the protoplanetary disk where the Solar System originated \citemain{OBrien2018}. However, it is unclear whether the water ice on these bodies primarily formed in e.g., the protoplanetary disk phase or is much older and originated from the parent molecular cloud \citemain{Cleeves2014}. This is because the bulk water reservoir is extremely difficult to detect and follow across the various phases of star and planet formation \citemain{Hogerheijde2011, Persson2016, Du2017, Notsu2019, Harsono2020, vanDishoeck2021}. If the water molecules that formed in the cloud survive the cloud collapse into a protostar with a surrounding envelope and subsequently the formation of a disk, then protoplanetary disks are injected with a rich chemical icy inventory as many complex organic molecules (COMs) have formation pathways in the ice \citemain[e.g., ][]{Watanabe2002, Garrod2006, Potapov2021}. This potential ice inheritance would set the initial conditions for the chemical inventory of protoplanetary disks and is crucial to interpreting observations of COMs in disks. Intriguingly, if the water ice survives the journey from cloud to disk to comets and possibly (exo-)planets, then the water seen across all these phases and possibly on Earth predates the formation of the star and our Sun, respectively \citemain{Cleeves2014}.  

A sensitive tracer to distinguish the inheritance and (partial) reset of water ice from pre-stellar cores to protoplanetary disks is the deuteration level of water. In case of reset, at least one of the chemical bonds between the atoms of the water molecule is destroyed. The deuteration is only enhanced above the D/H ratio of $2\times 10^{-5}$ in the interstellar medium (ISM) \citemain{Linsky2003} under specific conditions: below 25~K and where the density is sufficiently high for the freeze-out of atoms and molecules (few times $10^4$~cm$^{-3}$) \citemain[e.g.,][]{Watson1976, Ceccarelli2014}; therefore, water ice formed at these low temperatures will be rich in deuterated water isotopologues. On the contrary, reprocessing of this pristine material at temperatures higher than 500~K can lower the abundance of deuterated water isotopologues by destroying the molecules themselves \citemain{Owen2015, Furuya2016, Furuya2017}. The high level of singly deuterated water (HDO) seen in protostellar envelopes, disks, and comets has been presented as evidence of the inheritance of water \citemain[and references therein]{Ceccarelli2014, Tobin2023, Slavicinska2024}. However, the abundance of HDO can be as high as $\sim10^{-3}$ with respect to \ce{H2O} even if the majority of the water reservoir is reprocessed \citemain{Furuya2017}. 
Instead, doubly deuterated water (\ce{D2O}) is the most sensitive tool to distinguish inheritance from reset because it cannot reform efficiently after reprocessing. Yet, observations targeting this isotopologue are rare across all star- and planet formation phases because its abundance is expected to be low even if deuteration is enhanced because of the double deuteration and the low D/H of the ISM.

In particular, observations of doubly deuterated water are missing in the protoplanetary disk stage. The few existing observations probing water ice in disks cannot distinguish whether water is formed in the initial stages of star and planet formation or in situ due to a lack of detected isotopologues \citemain[e.g.,][]{Chiang2001, Sturm2023}. In addition, most protoplanetary disks are too cold to host a large and observable reservoir of gas-phase water. Tracers other than water isotopologues provide evidence for both inheritance and reprocessing depending on the disk and tracers observed. 
Inheritance is suggested by the detection of gas-phase methanol in three disks that are too warm for significant CO freeze-out \citemain{Booth2021HD100, vanderMarel2021_IRS48, Booth2023, Leemker2023, Leemker2024}. As methanol can only form efficiently when CO is frozen-out, the methanol ice in those disks must have formed before the envelope collapsed to form a disk and survived this process \citemain{Garrod2006, Geppert2006}. On the other hand, the deuteration seen in methanol and other complex organic molecules in the same disk as the one analyzed in this work, V883~Ori, is lower than expected based on a much younger protostellar envelope and a much older comet. This has been interpreted as an indication for some reprocessing of the ices in this disk \citemain{Yamato2024}, yet possibly \ce{CH3OH} is slightly more reprocessed than water due to its lower sublimation temperature \citemain{Minissale2022}. We present observations in the V883~Ori disk of doubly-deuterated water \ce{D2O}, the most sensitive tracer to distinguish inheritance from reprocessing to shed light on the origin of ices in disks, in particular the origin of water ice.

\section{Results}
We detect the \ce{p-D2O} $1_{1,0}-1_{0,1}$ transition at 316.7998~GHz in the disk surrounding the young, outbursting V883~Ori star with the Atacama Large Millimeter/submillimeter Array (ALMA). This star is located at 400~pc (1300 ly) in the Orion molecular cloud \citemain{Kounkel2017} and has a source velocity of 4.3~km~s$^{-1}$ \citemain{Cieza2016, vantHoff2018}. Freshly sublimated ices are seen in this disk due to the heating from the outbursting star allowing for a unique look into the bulk water reservoir \citemain{vantHoff2018, Lee2019, Tobin2023}. We compare our \ce{D2O} detection with the previously observed HDO and H$_2^{18}$O emission in this disk \citemain{Tobin2023}.  

\begin{figure}[t]
    \centering
    \includegraphics[width=1\linewidth]{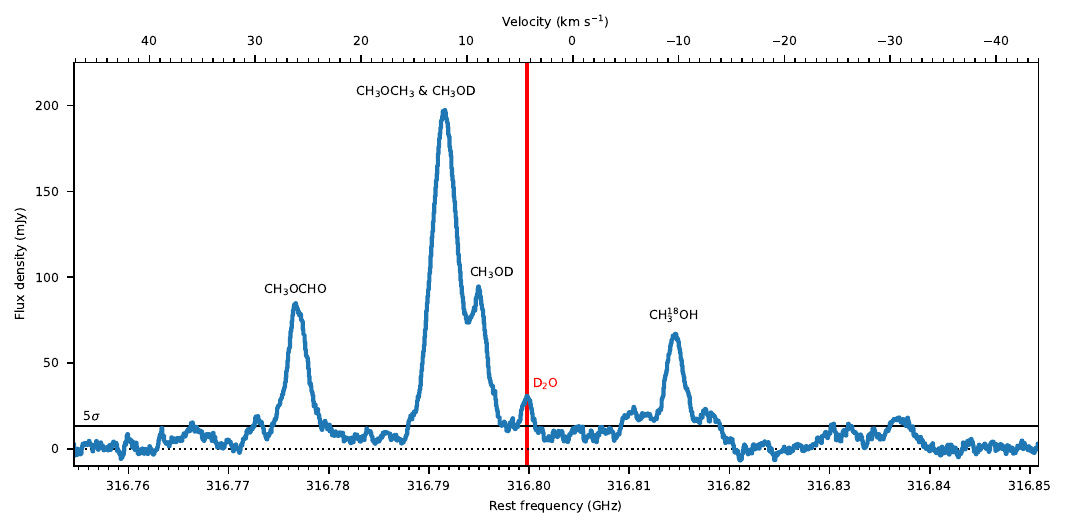}
    \caption{Integrated spectrum of \ce{D2O} in the V883~Ori disk. The spectrum is extracted from an elliptical region with a $0\farcs4$ semi-major axis and a $0.4"\times \cos(i)$ semi-minor axis centered on the continuum peak, with $i$ the disk inclination. The pixels in the image cube were shifted by the projected Keplerian velocity at that location in the disk before extracting the spectrum to correct for the projected Keplerian rotation of the disk and decrease line blending of \ce{D2O}, indicated with the red vertical line, with the neighboring \ce{CH3OD} and \ce{CH3OCH3} lines. The solid horizontal line indicates the 5$\sigma$ noise level measured in off-source regions. } \label{fig:spec}
\end{figure}

The \ce{D2O} line observed in this program is blended with emission from neighboring transitions of deuterated methanol (\ce{CH3OD}) at 316.7916~GHz ($3_{2, 1}-3_{-1, 3}$ A) and 316.7951~GHz ($7_{0, 7}- 6_{0, 6}$ A) and two dimethyl ether (\ce{CH3OCH3}) transitions at 316.7915~GHz ($22_{6,16}-22_{5,17}$ EE) and 316.7925~GHz ($22_{6,16}-22_{5,17}$ AA), respectively, similar to observations of younger Class~0 objects \citemain{Jensen2021b}. 
To separate the \ce{D2O} from these COMs, each pixel in the image cube is shifted by the projected Keplerian velocity associated to that pixel in order to remove the rotation component of the line velocity, before extracting the disk-integrated spectrum in an elliptical region with a $0\farcs4$ semi-major axis presented in Figure~\ref{fig:spec}. The non-shifted version is presented in Extended Data Figure~1.
The \ce{D2O} line is detected at a peak signal-to-noise ratio of 11 in the shifted spectrum, where the noise is estimated using the rms of 520 independent shifted spectra extracted from a 19" square in the cube without primary beam correction, excluding the inner $2\farcs4$ square to avoid possible contamination with extended line emission. In addition, the channel maps show significant emission across multiple channels that can only be attributed to the \ce{D2O} molecule and not to any neighboring lines of complex organic molecules. . 
An overview of the \ce{D2O}, HDO, and H$_2^{18}$O line properties is presented in Table~\ref{tab:CDMS_JPL}. The line flux is measured both from the channel maps presented in Extended Data Figure~2, 3, and 4 and from the shifted spectra presented in Extended Data Figure~5, 6, 7, and 8. An overview of these line fluxes is presented in Table~\ref{tab:fluxes}, where the reimaged HDO and H$_2^{18}$O emission originally presented in \citemain{Tobin2023} are included for consistency. The flux measured using the Keplerian masks is considered to be the fiducial flux (see Sect.~\ref{sec:flux} for details).
 
As the spatial distribution of all three water isotopologues (\ce{D2O}, HDO, and H$_2^{18}$O) is similar (see Figure~\ref{fig:D2O_channel_maps_cf_HDO} for the channel maps of \ce{D2O} and HDO and \cite{Tobin2023} for a comparison of HDO and H$_2^{18}$O), we use the excitation temperature of $199\pm 42$~K of the HDO molecule \citemain{Tobin2023}, the only water isotopologue with two observed transitions, for all the water isotopologues. The effect of the uncertainty on this excitation temperature is discussed in Supplementary Sect.~3.1 as one of the HDO lines used for this measurement is heavily blended with an emission line of a COM, possibly affecting the derived excitation temperature. The line fluxes are measured using Keplerian masks that separate the water isotopologue emission from that of the neighbouring COMs in both Right Ascension, Declination, and velocity. These line fluxes of the \ce{D2O}, HDO, and H$_2^{18}$O transitions are then converted to column densities of $(4.2 \pm 1.2)\times 10^{13}$~cm$^{-2}$, $(49.5 \pm 6.9) \times 10^{14}$~cm$^{-2}$, and $(23.7 \pm 5.7) \times 10^{14}$~cm$^{-2}$ assuming an emitting area equal to the elliptical region used to extract the spectrum and the flux integrated over the velocity channels where all three water isotopologues can be separated from other emission lines (see also Sect.~\ref{sec:flux}, \ref{sec:rot_dia}, Supplementary Sect.~3, and Table~\ref{tab:N}). As no strong isotope selective effects are expected between \ce{H2O} and H$_2^{18}$O, the H$_2^{18}$O column density is scaled to a total \ce{H2O} column density of $(13.3\pm 3.2) \times 10^{17}$~cm$^{-2}$ using the typical \ce{^16O/^18O} ratio of $560 \pm 25$ in the ISM \citemain{Wilson1994}, consistent with the lower limit derived in the HL~Tau disk of $N($H$_2^{16}$O$)/N($H$_2^{18}$O$) >40$ \citemain{Facchini2024}. These column densities translate to \ce{D2O}/\ce{H2O} = $(3.2 \pm 1.0) \times 10^{-5}$ and (\ce{D2O}/HDO) / (HDO/\ce{H2O}) = $2.3 \pm 1.0$.

\section{Discussion and conclusions}\label{sec:disc_concl}
\subsection{Water isotopologue ratios along the water trail}
Both the \ce{D2O}/\ce{H2O} and the (\ce{D2O}/HDO) / (HDO/\ce{H2O}) ratios in the V883 Ori disk are similar to the observed values of \ce{D2O}/\ce{H2O} = $(2-9)\times 10^{-5}$ and (\ce{D2O}/HDO) / (HDO/\ce{H2O}) = $2.2-7.1$ in younger Class~0 objects \citemain{Coutens2014, Jensen2019, Jensen2021b}, which are expected to trace pristine ices. In addition, the \ce{D2O}/\ce{H2O} ratio in the V883~Ori disk is similar to that of $(1.9 \pm 1.0)\times 10^{-5}$ in the comet 67~P \citemain{Altwegg2015, Altwegg2017}, see Figure~\ref{fig:inheritance_vs_reset}. This strongly indicates that the ice in the V883~Ori disk is inherited from the cold molecular cloud that collapsed to form a Class~0 protostar with protostellar envelope and subsequently an embedded disk, and potentially connects to the cometary phase.

The \ce{D2O} abundance in the V883~Ori disk is not expected to be greatly altered by the outburst as modelling of HDO and \ce{H2O} in a 1D evolving disk around an outbursting star has shown that the outburst only affects the water deuteration in the inner 1-3~au \citemain{Owen2015}. The emission of the water isotopologues in the V883~Ori disk is only seen much further out at radii larger than 40~au, as optically thick dust hides the line emission inside that radius \citemain{Cieza2016, Tobin2023}.

The colored background in Figure~\ref{fig:inheritance_vs_reset} indicates the range of \ce{D2O}/\ce{H2O} ratios consistent with inheritance (blue) and reset (red) based on chemical modelling of gas and ice in a collapsing core by \citemain{Furuya2017}. These simulations follow gas parcels as the rotating core collapses from the inside out and forms a disk and envelope. These stream lines are post processed with a dedicated chemical network modelling the deuterium chemistry in the gas, a chemically active surface layer of ice interacting with the gas, and a chemically inert bulk ice phase. As these stream lines trace a range in physical conditions, also the water isotopologue ratios show a range of possible values for either inheritance (all stream lines where $\lesssim 10$\% of the water ice is destroyed) or reset (all stream lines where $\gtrsim 70$\% of the water ice is destroyed). The V883~Ori datapoint coincides with the peak of the model prediction histogram for inheritance well within $1\sigma$, further supporting that the material in this disk is inherited. 

The observed (\ce{D2O}/HDO) / (HDO/\ce{H2O}) ratio is consistent between Class~0 objects and the V883~Ori disk and are all a factor of $\sim 40-140$ higher than the most likely value of $\sim 5\times 10^{-2}$ in the case the material is reset (see Extended Data Figure~9). The (\ce{D2O}/HDO) / (HDO/\ce{H2O}) ratio predicted in \citemain{Furuya2017} for inheritance is 10, which is a factor up to $\sim 5$ higher than that observed. Nonetheless, in contrast to the values for reset, the initial (\ce{D2O}/HDO) / (HDO/\ce{H2O}) can vary between $\sim 0.8$ and $\sim 19$ due to variations in the initial conditions of the models \citemain{Furuya2016} (see the Supplementary Information for details). The similarity in the observed ratio between the V883~Ori disk and the Class~0 objects strongly suggests that material is inherited. 

As all three analyzed transitions lie at different frequencies of 316.7998~GHz (\ce{D2O}), 225.8967~GHz (HDO), and 203.4075~GHz (H$_2^{18}$O), frequency dependent radiative transfer effects could	affect the observed \ce{D2O}/\ce{H2O} and the (\ce{D2O}/HDO) / (HDO/\ce{H2O}) ratios. In particular, the dust inside 40~au is optically thick \citemain{Cieza2016, Houge2024} absorbing the emission of COMs and other molecules whose emission partly originates at altitudes comparable to the optically thick layer of the continuum emission \citemain{Isella2016, Weaver2018, Lee2019, Tobin2023, Yamato2024}. As the continuum optical depth increases with frequency, more \ce{D2O} emission could be hidden than HDO and H$_2^{18}$O due to the higher frequency of the \ce{D2O} line. Therefore, if optically thick dust is affecting the lines differentially, the measured ratios will be driven further into the regime consistent with inheritance. Other systematics that could affect the measured \ce{D2O}/\ce{H2O} and (\ce{D2O}/HDO) / (HDO/\ce{H2O}) ratios such as a lower excitation temperature are discussed in Supplementary Sect.~3.

\subsection{Deuteration of complex organic molecules}

The high abundance of \ce{D2O} with respect to \ce{H2O} in the V883~Ori disk demonstrates that water in this disk is likely inherited. This is further supported by the high gas-phase HDO/\ce{H2O} ratio seen across protostellar envelopes, disks, and comets \citemain[and references therein]{Tobin2023} and the high HDO/\ce{H2O} ratio seen in the ice in a single low mass protostar and two massive protostars \citemain[]{Slavicinska2024, Slavicinska2025}. However, the abundance of \ce{CH2DOH}, \ce{CH3CDO}, \ce{CH3OCDO}, and \ce{CH2DOCHO} compared to their main isotopologues is measured to be lower in the V883 Ori disk than in the younger hot corino IRAS~16296~B and the comet 67P. This lower level of deuteration in the V883~Ori disk has been interpreted as an indication for some reprocessing of COMs \citemain{Yamato2024, Jeong2024}. 

COMs are traditionally thought to form in CO rich ice but there are indications in the V883~Ori disk that COMs possibly partially formed in a water-rich ice matrix (\ce{H2O}) \citemain{Jeong2024}. The contribution from COMs formed in the CO-rich ice matrix is expected to have a higher D/H ratio as deuteration is efficient at the temperature where CO freezes-out. The initial deuteration levels of these COMs likely follows the \ce{D2O}/HDO ratio as \ce{D2O} and HDO are formed at the same time. The levels of deuteration seen in a \ce{CH2DOH}/\ce{CH3OH} ratio of $1.97 \times 10^{-2}$ \citemain{Lee2019, Jeong2024} and the low upper limits on the deuteration in \ce{CH2DOH}, \ce{CH3CDO}, \ce{CH3OCDO}, and \ce{CH2DOCHO} of $\lesssim (5.7-20) \times 10^{-3}$ \citemain{Yamato2024} are consistent with the \ce{D2O}/HDO ratio of $(8.5 \pm 2.8) \times 10^{-3}$ in this disk. Yet, chemical reactions where H and D are exchanged or abstracted can alter the D/H ratio in COMs complicating the analysis \citemain[e.g.,][]{Taquet2012}.
Even though COMs are likely to at least partially form in the CO-rich ice matrix, the contribution of COMs formed within the water ice matrix will have a low deuteration level due to the higher gas temperatures at the time of formation making deuteration inefficient.

\subsection{Summary}

The inheritance of water from the ice on dust grains in the cloud before stars are born to the disk and subsequently to comets connects the major steps in the formation of planets as comets likely form from the same reservoir of material as the planets. Our result demonstrates that pristine water that formed at the earliest phases of star and planet formation is available in protoplanetary disks, a phase where potential hints of embedded planets are frequently seen in the form of e.g., dust substructures \citemain[e.g.,][]{Bae2023}. Even though the journey of water from the disk to Earth is still debated and the deuteration in Earth's oceans is lower than HDO/\ce{H2O} ratio in the V883 Ori~disk, \citemain{OBrien2018, Tobin2023}, the deuteration of Earth's oceans is enhanced compared to the ISM by an order of magnitude \citemain{Hagemann1970, DeLaeter2003, Linsky2003}. 
In addition, the \ce{D2O}/\ce{H2O} ratio is similar across Class~0 objects, the V883~Ori disk, and a comet. Together with the enhanced deuteration seen in HDO across all these four phases, this suggests that the trail of inherited water does not stop at the cometary phase but potentially continues to the water present on exoplanets formed in water ice rich protoplanetary disks.

\section{Methods}\label{sec:methods}

\subsection{Self-calibration}

We observed the V883~Ori disk in ALMA band~7 for 2.4 hours on source distributed over three execution blocks (EBs; Supplementary Data Table~1; 2023.1.00588.S; PI: M. Leemker). These data targeted the \ce{D2O} $1_{1,0} -1_{0,1}$ transition at 316.7998~GHz at a spectral resolution of 61~kHz (58~m~s$^{-1}$) and a total bandwidth of 117.2~MHz. A continuum spectral window covering 1.875~GHz of bandwidth at a spectral resolution of 1.1~km~s$^{-1}$ was centered at 315~GHz. In addition, 6 spectral windows centered at the 
\ce{^13CN} $N=3-2$, $J=5/2-3/2$, $F_1=2-1$, $F=3-2$ (325.943~GHz),
\ce{N2O} $13_{0,0}-12_{0,0}$ (326.556~GHz),
\ce{N2O} $13_{-1,2}-12_{1,2}$ (326.685~GHz),
\ce{CH3OCH3} $18_{1,18}-17_{0,17}$ EE (326.931~GHz),
\ce{NO2} $22_{1,21}-22{0,22}, J=45/2-45/2, F=43/2-43/2$ (328.097~GHz), and
\ce{NO2} $22_{1,21}-22_{0,22}, J=45/2-45/2, F=47/2-47/2$ (328.131~GHz)  
transitions were included with a spectral resolution of 122~kHz ($\sim 130$~m~s$^{-1}$) and a bandwidth of 58.6~MHz. Finally, the 
\ce{CH3OCH3} $10_{3,8}-9_{2,7}$ EE transition at 328.857~GHz was targeted at the same spectral resolution but a bandwidth of 117.2~MHz.

The data were self-calibrated using CASA version 6.5.4 \citemain{McMullin2007} following the analysis used in the exoALMA large program outlined in \citemain{LoomisexoALMA}, with several routines by \citemain{Andrews2018, Czekala2021}. A pseudo-continuum Measurement Set was created for each EB after carefully flagging all spectral regions exhibiting line emission, and then averaging all the data in individual channels for each spectral window. Since this is a very line-rich disk, the total continuum bandwidth used for the self-calibration is reduced to 180~MHz. Before aligning and then combining the three EBs, one round of phase self-calibration over a solution interval equal to the length of a single EB (1 hour and 19 minutes) was performed combining all scans, and both polarizations. A \texttt{CLEAN} model was used for the self-calibration, as constructed over an elliptical mask of $0\farcs6$ in semi-major axis, and a Position Angle (PA) and semi-minor axis computed from the PA and inclination of the V883 disk of 32$\degree$ and 38.3$\degree$, respectively \citemain{Cieza2016}. We employed Briggs weighting with \texttt{robust}$=0.5$. The model was created by cleaning the emission in all line-free channels down to a conservative threshold of $6\sigma$. The peak signal-to-noise ratio in the images of the individual EBs improved between 76 and 160\%. The continuum observed in the third EB had the highest signal to noise ratio of all EBs and therefore was used as a reference for the alignment where the phase angle and the amplitude differences are minimized between the first EBs and the third EB \citemain{LoomisexoALMA}. 

After combining all execution blocks, seven rounds of phase-only self-calibration were performed with solution intervals with the length of a single EB, 360, 120, 60, 20, 10, and 6 seconds. The imaging followed the same mask and weighting as for the individual EBs before concatenation. In the first round of phase self-calibration, separate solutions were found for both polarizations, whereas in all subsequent rounds a single solution was obtained for both polarizations. In addition, all scans were combined during the first three rounds as the solution interval exceeded the scan length. 

Comparing the amplitude of the continuum visibilities between the three execution blocks showed that the data taken on December 24th 2023 had an 8.6\% higher amplitude across all baselines up to 600 k$\lambda$ where the signal-to-noise ratio drops compared to the execution block with the highest signal-to-noise-ratio. Therefore, we rescaled the flux before self-calibration in this execution block and concatenated this to the data of the other two EBs before self-calibration. Subsequently, these concatenated data were self-calibrated with seven rounds of phase-only self-calibration and a single round of phase and amplitude self-calibration with separate solutions for both polarizations and a solution interval of the length of a single EB. The model for the amplitude self-calibration was created by cleaning the line-free channels down to a $1\sigma$ threshold to capture as much emission as possible in the model. Solutions deviating by more than $\sim$20\% in amplitude were flagged. The phase self-calibration improved the peak signal-to-noise ratio of the data by a factor of 186\%, from a peak signal-to-noise ratio of 495 to one of 1457. The amplitude and phase self-calibration improved the peak signal-to-noise ratio by 3\% (see Supplementary Figure~1 for the final continuum image). The peak continuum intensity increases by 15\% after phase-only self-calibration due to the improved phase coherence after phase-only self-calibration. The peak continuum intensity changes by less than 0.5\% after phase and amplitude self-calibration. The solutions were then applied to the full dataset, following the same logical order as for the pseudo continuum Measurement Set.

\subsection{Continuum subtraction}
As many lines of COMs are detected in this dataset, too few line-free channels were available to subtract the continuum emission in the $uv$-plane using the standard CASA task \texttt{uvcontsub}. Instead, the continuum in the \ce{D2O} spectral window was subtracted in each EB separately. First, a mtmfs image up to and including first order (\texttt{nterms} $=2$) was created using the line free channels in the \ce{D2O} spectral window. The resulting model for the continuum emission was added to the measurement set using \texttt{ft}. This model was then used to subtract the continuum using the CASA task \texttt{uvsub}. The resulting MS tables for all three EBs were then combined to make the final continuum subtracted dataset for \ce{D2O}. 

\subsection{Imaging}
The \ce{D2O} line itself was imaged using the CASA task \texttt{tclean} with Briggs weighting and a \texttt{robust} parameter of 0, which provides the best balance between spatial resolution and noise, resulting in a $0\farcs 24 \times 0\farcs 19\ (89\degree)$ beam. The spectrum around the \ce{D2O} emission is presented in Extended Data Figure~1 and the \ce{D2O} channel maps are presented in Figure~\ref{fig:D2O_channel_maps_cf_HDO}. The relatively small beam allows for better deblending of the lines while still detecting the line. The image is cleaned down to $1\sigma$ to capture all the flux in the model and make the final measurement independent of any non-Gaussian features in the beam at the cost of potential imaging artifacts \citemain{Jorsater1995, Czekala2021}. The channel spacing for the final image is 400~m~s$^{-1}$ to match that of the HDO and H$_2^{18}$O data. The resulting cubes are used to produce channel maps and measure the flux using a Keplerian mask. Only when the data cubes are shifted with the projected Keplerian velocity associated to each pixel, the native spectral resolution of the data of 58, 162, and 180~m~s$^{-1}$ for \ce{D2O}, HDO, and H$_2^{18}$O, respectively, is used to minimize artifacts due to the finite spectral resolution of the data.

To accurately measure the (\ce{D2O}/HDO) / (HDO/\ce{H2O}) and \ce{D2O}/\ce{H2O} ratio in this disk, we re-imaged the HDO and H$_2^{18}$O data presented in \citemain{Tobin2023} (2021.1.00186.S; PI: J. J. Tobin). In this work, these data are cleaned down to $1\sigma$, similar to the \ce{D2O} line, and allow for a uniform analysis of the line fluxes and migitate any effects of the non-Gaussian beam shape on the measured line flux \citemain[e.g.,][]{Czekala2021}. The resulting beam sizes are $0\farcs13\times  0\farcs11 (-72\degree)$ and $0\farcs10\times 0\farcs08 (-78\degree)$ for the H$_2^{18}$O and the HDO when cleaned with a \texttt{robust} parameter of 1.0 and 2.0, respectively, to provide the best balance between the signal-to-noise ratio and a small beam to separate the lines from neighbouring COMs. The resulting channel maps for HDO and H$_2^{18}$O are presented in Extended Data Figure~3 and Extended Data Figure~4, respectively.

\subsection{Flux measurements} \label{sec:flux}
The transitions of all three water isotopologues discussed in this paper are blended with lines from neighboring COMs. To separate these lines and accurately measure the line flux, two methods are employed: Keplerian masking in the channel maps and spectral shifting of the spectrum. In the former case, the lines are separated in the image plane by computing the region in Right Ascension and Declination where a molecule is expected to emit based on the Keplerian rotation of the disk and its orientation on the sky. As this region shifts for each velocity channel, lines of neighbouring COMs can be separated from the water isotopologue emission. The Keplerian masks are constructed using the package by \citemain{Teague2020} following \citemain{Tobin2023} using a distance of 400~pc \citemain{Kounkel2017}, a stellar mass of 1.3~M$_{\odot}$, a disk inclination of 38.3$\degree$, and a position angle of 32$\degree$ \citemain{Cieza2016} and an emitting region extending from 40~au to 120~au tracing $z/r=0.4$. The line width is assumed to follow 350~m~s$^{-1} \times \sqrt{40~\mathrm{au}/r}$ and the mask is shifted by 4.3~km~s$^{-1}$ to match the systemic velocity of V883~Ori \citemain{Cieza2016, vantHoff2018}. These parameters are identical between all three analyzed water isotopologues. Finally, the mask is convolved with a Gaussian 0.5 (\ce{D2O}), 1.25 (HDO), and 0.75 (H$_2^{18}$O) times larger than the beam of the respective observations. This parameter is chosen after visual inspection of the data. The factor varies slightly from line to line to match the spatial extent of the emission observed in the channel maps and to match the spatial resolution of the resulting masks to have a $\sim 0.1-0.12"$ major axis and a $0.08-0.1"$ minor axis across all lines. The effect of neighboring COMs is investigated using an identical mask but then shifted to the frequency of those lines. 

In general the emission of the \ce{D2O}, HDO 225~GHz, and the H$_2^{18}$O transition can be well separated in most channels using this method. To measure the disk integrated flux, only the channels between 1.2 and 6.4~km~s$^{-1}$ are used as only in these channels the water emission can be separated from the COMs for all three isotopologues. The uncertainties on the disk integrated line fluxes are estimated as the rms of the flux measured in masks shifted to eight different positions $1\farcs4$ from the original location and to five different velocities resulting in 40 unique, non-overlapping locations. The HDO 241~GHz line on the other hand is too severely blended and is therefore not considered in this analysis.

The second method to deblend the lines is through spectral shifting. Similar to the Keplerian masking, this method utilizes the Keplerian rotation of the disk. Each pixel is shifted to correct for the projected Keplerian velocity expected at this location in the disk. The lines in the spectrum extracted from such a shifted cube are much narrower and single peaked than the double peaked line profiles in the non-shifted cube. As the lines are only seen over a small radial range, the emission is assumed to be originating from the disk midplane. 

The resulting line profiles are not necessarily Gaussian. Therefore, the line profile of each COM is reconstructed from the data directly, see Extended Data Figures~5, 6, and 7. The non-blended side of COM emission line is mirrored in the line center creating a symmetric line profile that is then subtracted from the shifted spectrum. This procedure is repeated for all bright lines of COMs around the water lines. The \ce{D2O} and H$_2^{18}$O line were still blended by some remaining COM emission. Therefore, a spectral template for the \ce{D2O} and H$_2^{18}$O themselves were created using the clean side of the line after the COM emission was subtracted (see the cyan line in the bottom panel of Extended Data Figures~5 and 7). For the H$_2^{18}$O, the contribution of the contaminating shoulder is negligible at $<1\sigma$, whereas for the \ce{D2O} the \ce{CH3OD} contributed 8~mJy~km~s$^{-1}$ to the total \ce{D2O} flux. The region used to create the line profile of the COMs is indicated with the blue shaded region and that used to measure the water isotopologue line flux is indicated in red. 

Extended Data Figure~8 presents a spectrum of the H$_2^{18}$O line but then extracted from a somewhat smaller region of $0\farcs35 \times 0\farcs27$ (semi-axes). The noise is estimated from 720 independent spectra covering a 19" region in the image plane while excluding the central $2\farcs1$ square in the non-primary beam corrected cube. As less noise is added to this spectrum than in Extended Data Figure~7, it clearly shows an emission peak at the expected frequency for the H$_2^{18}$O line, showing that this line is detected in this dataset. This spectrum is not used for any of calculations to avoid a bias due to the smaller region used to extract this spectrum compared to that of the \ce{D2O} and HDO lines.

An overview of the measured fluxes is presented in Table~\ref{tab:fluxes}, where for the \ce{D2O} and H$_2^{18}$O emission, we report the flux measured by the spectral template centered on those lines to account for the remaining shoulder of COM emission in the spectrum. The reported uncertainties on the line fluxes do include the statistical uncertainty but not the 5\% absolute flux calibration uncertainty for ALMA Band~5 and 6 (HDO and H$_2^{18}$O) and 10\% for ALMA Band~7 (\ce{D2O}). The absolute flux calibration uncertainty is included for all line ratios together with the uncertainty on the \ce{^16O}/\ce{^18O} ratio of $560 \pm 25$ in the ISM \citemain{Wilson1994}. The uncertainty on the \ce{^16O}/\ce{^18O} ratio in the ISM is negligible for the resulting \ce{H2O} column density and water isotopologue ratios. 

The two methods to measure the line fluxes are consistent within $\leq 3\sigma$ for all three water isotopologues. This difference stems from the limited velocity range used for the Keplerian masking method. In this case only the channels between 1.2 and 6.4~km~s$^{-1}$ are used as all three water isotopologues can be separated from the neighbouring COMs in these channels, whereas emission in some isotopologues is seen up to 0.8 and 8.4~km~s$^{-1}$. 
As some flux from the inner disk region may be lost due to the spectral shifting of the data as a result of beam smearing and the main goal of this work is to obtain ratios of water isotoplogues, the fluxes measured using the Keplerian masks are used for the final analysis presented in the main text.

\subsection{Column density estimates} \label{sec:rot_dia}
\begin{figure*}[t]
    \centering
    \includegraphics[width=\linewidth]{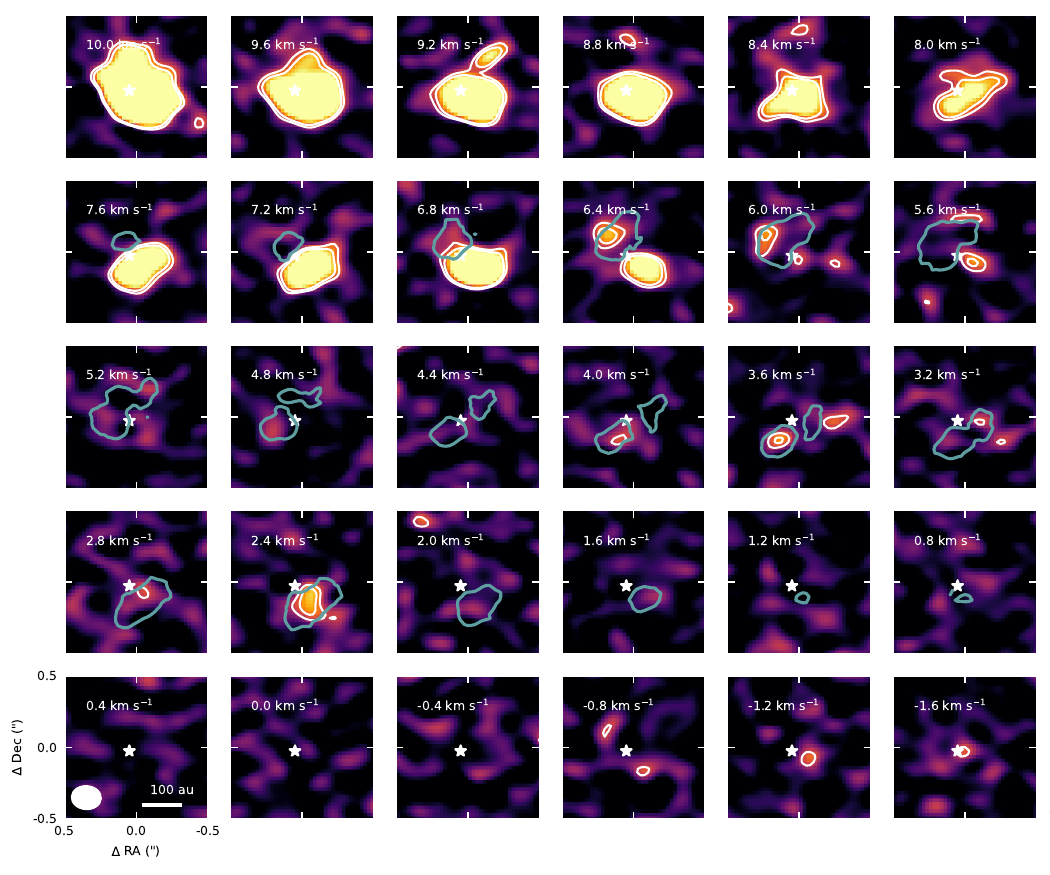}
    \caption{Channel maps of the \ce{D2O} emission compared to that of HDO. The grey line indicates the emission above $3\sigma$ attributed to the HDO line at 225~GHz. The white contours indicate the 3 and 4$\sigma$ level for the emission of \ce{D2O} and the neighbouring COMs. } \label{fig:D2O_channel_maps_cf_HDO}
\end{figure*}

\begin{figure}[t]
    \centering
    \includegraphics{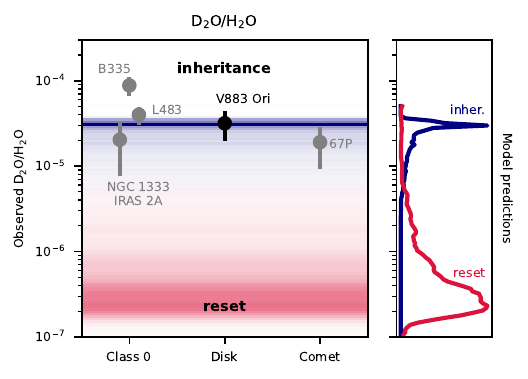} 
    \caption{The \ce{D2O}/\ce{H2O} ratio across different stages of star and planet formation. The measurements in the V883 Ori disk are presented in black and those in the Class~0 objects NGC~1333~IRAS~2A, B335, L483, and the 67P comet in grey \protect\citemain{Coutens2014, Altwegg2015, Altwegg2017, Jensen2019, Jensen2021b}. The latter are computed as \ce{D2O}/HDO $\times$ HDO/\ce{H2O}. The errorbars represent the 1$\sigma$ uncertainty (s.d.) on the measured column density ratio in each source. The colored background and the histograms on the side, each normalized to the peak number of fluid parcels, indicate the expected water isotopologue ratios for inheritance where $\lesssim 10$\% of the \ce{H2O} ice is destroyed (blue) or reset where $\gtrsim70$\% of the \ce{H2O} is expected to be destroyed through photodissociation and photodesorption in a model of a collapsing core (red; \protect\citemain{Furuya2017}). The red histogram is smoothed using a Savitzky-Golay filter with a window of 10 and an order of 3. }  \label{fig:inheritance_vs_reset}
\end{figure}

In order to compute the ratios of water isotopologues, their line fluxes need to be converted to column densities using the properties of the emission lines listed in Table~\ref{tab:CDMS_JPL}. In this work, optically thin emission and local thermodynamical equilibrium (LTE) are assumed for this conversion. These assumptions are further discussed in Supplementary Sect.~3.2 and 3.4, respectively.  
The column density of the upper energy level can be determined without any knowledge on the rotational temperature \citemain{Goldsmith1999, Loomis2018}:
\begin{align}
    N_u = \frac{4\pi F_{\nu}\Delta V}{A_{ul}\Omega hc},
\end{align}
with $\int F_{\nu} dV \approx F_{\nu}\Delta V$ the integrated flux, $A_{ul}$ the Einstein-A coefficient of the transition, $\Omega$ the emitting region in steradian, $h$ the Planck constant, and $c$ the speed of light. The water snowline, the midplane region where water sublimates is possibly at 40~au but most likely around $75-120$~au based on gas-phase measurements \citemain{Cieza2016, Tobin2023}, \citemain{vantHoff2018}, \citemain{Leemker2021}, \citemain{Wang2025}. In addition, gas-phase water traced by HDO and H$_2^{18}$O is seen out to 160~au ($0\farcs4$) \citemain{Tobin2023} and as the spatial distribution of the \ce{D2O}, HDO, and H$_2^{18}$O water isotopologues is similar (see Figure~\ref{fig:D2O_channel_maps_cf_HDO}, Extended Data Figures~2, 3, and 4) an elliptical emitting region with that radius is used. The minor axis of this elliptical region is set to $0\farcs4 \times \cos(i) = 0\farcs 31$, with $i$ the disk inclination to trace a circular region with a radius of 160~au in the disk frame.

The column density of the upper energy level is then converted to the total column density using the partition function $Q$ at the rotational temperature $T_{\rm rot}$ and the degeneracy $g_u$ and the upper energy level $E_u$ of the line:
\begin{align}
    N_{\rm tot} = \frac{N_u Q(T_{\rm rot})}{g_u}\exp{\left (\frac{E_u}{k_{\rm B}T_{\rm rot}} \right )},
\end{align}
with $k_{\rm B}$ the Boltzmann constant. If multiple lines of the same molecule are observed, this equation can be solved for each line to find the excitation temperature of the molecule. In this work, the excitation temperature derived by \citemain{Tobin2023} from two HDO transitions with different upper energy levels is used for all analyzed lines as these water isotopologues emit from the same disk region.

For the \ce{D2O} and H$_2^{18}$O molecules that contain two deuterium or two hydrogen atoms, the molecules can be either in an ortho or in a para-state. The transitions used in this work are both para-states. The column density of \ce{p-D2O} and p-H$_2^{18}$O are converted to the total column density using a thermalized ortho-to-para ratio of 2 and 3, respectively \citemain{Vastel2010, Hama2016}. A summary of the derived column densities are presented in Table~\ref{tab:N}.

\section{Declarations}

\subsection{Data availability}
The ALMA data from ALMA programs \#2021.1.00186.S and \#2023.1.00588.S are publicly available on the ALMA archive on the ALMA archive at \url{https://almascience.eso.org/aq/}.

\subsection{Code availability}
The code used for this work is based on publicly available python packages. In addition, the code itself is available upon reasonable request to M.L.. 

\subsection{Acknowledgements}
We would like to thank the referees for the constructive comments. M.L. would like to thank Ewine van Dishoeck, Milou Temmink, and Shota Notsu for the useful discussions and acknowledges assistance from Allegro, the European ALMA Regional Center node in the Netherlands.
M.L. and S.F. are funded by the European Union (ERC, UNVEIL, 101076613). Views and opinions expressed are however those of the author(s) only and do not necessarily reflect those of the European Union or the European Research Council. Neither the European Union nor the granting authority can be held responsible for them. 
S.F. acknowledges financial contribution from PRIN-MUR 2022YP5ACE.
P.C. acknowledges support by the ANID BASAL project FB210003.
A.S.B. is supported by a Clay Postdoctoral Fellowship from the Smithsonian Astrophysical Observatory.
This paper makes use of the following ALMA data: ADS/JAO.ALMA\#2021.1.00186.S and \#2023.1.00588.S. ALMA is a partnership of ESO (representing its member states), NSF (USA) and NINS (Japan), together with NRC (Canada), MOST and ASIAA (Taiwan), and KASI (Republic of Korea), in cooperation with the Republic of Chile. The Joint ALMA Observatory is operated by ESO, AUI/NRAO and NAOJ. The National Radio Astronomy Observatory and Green Bank Observatory are facilities of the U.S. National Science Foundation operated under cooperative agreement by Associated Universities, Inc.

\subsection{Author contributions}
M.L. wrote the majority of the text of the manuscript and analyzed the data. M.L., P.C., J.J.T., and S.F. self-calibrated and imaged the ALMA data that was obtained by proposals led by M.L. and J.J.T. All co-authors provided input on the manuscript.

\subsection{Competing interests}
The authors declare no competing interests. 

\newpage

\FloatBarrier

\begin{table}
    \caption{Properties of the transitions analyzed in this work.}\label{tab:CDMS_JPL}
    \begin{tabular}{ccccccccc} 
        \toprule
        Molecule    & Transition & frequency           & $\log(A_{ul})$ & $E_u$ & $g_u$ & $Q$   & $Q$ & $Q$ \\
                    &  &  (GHz)  & $(\mathrm{s^{-1}})$ & (K)            &       & 300 K & 150 K & 75 K \\
        \midrule
        \ce{D2O}    & $1_{1,0}-1_{0,1}$ &  316.79981 & -3.20 & 32.6 & 3 & 349.6 & 123.7 & 44.4 \\%& ...\footnotemark[1] \\
        HDO         & $3_{1,2}-2_{2,1}$ & 225.89672 & -4.88 & 167.6 & 7 & 146.9 & 52.3 & 18.9 \\%& ... \\
        H$_2^{18}$O & $3_{1,3}-2_{2,0}$ & 203.40752& -5.31 & 203.7 & 7 & 179.6 & 64.2 & 23.4 \\%& ...\footnotemark[2] \\
        \botrule
    \end{tabular}
    \footnotetext{The \ce{D2O} line constants are taken from the CDMS database \citemain{Messer1984, Matsushima2001, Muller2001, Muller2005, Bunken2007, Endres2016} and those for HDO and H$_2^{18}$O are taken from the JPL database \citemain{Pickett1998, DeLucia1972, Messer1984}. The line constants for \ce{D2O} and H$_2^{18}$O use an ortho-to-para ratio of 2 and 3, respectively. }
    %	\footnotetext[1]{...}
\end{table}

\begin{table}
    \caption{Disk integrated line fluxes excluding emission from neighboring complex organics.}\label{tab:fluxes}
    \begin{tabular}{cccc} 
        \toprule
        Molecule & Transition & Line flux (mJy~km~s$^{-1}$) & Line flux (mJy~km~s$^{-1}$) \\
        &  & Keplerian mask & Shifted spectrum \\
        \midrule
        \ce{D2O} & $1_{1,0}-1_{0,1}$ & $53 \pm 4 $ & $61 \pm 4$ \\%& ...\footnotemark[1] \\
        HDO & $3_{1,2}-2_{2,1}$  & $358 \pm 14$ & $408 \pm 12$ \\%& ... \\
        H$_2^{18}$O & $3_{1,3}-2_{2,0}$  & $43 \pm 9$ & $58 \pm 10$ \\%& ...\footnotemark[2] \\
        \botrule
    \end{tabular}
    \footnotetext{The HDO and H$_2^{18}$O line fluxes are a reanalysis of those originally presented in \citemain{Tobin2023}. The uncertainty on the flux does not include the absolute flux calibration uncertainty of ALMA. The line fluxes are measured between 1.2 and 6.4~km~s$^{-1}$ for the Keplerian masking method.}
    %	\footnotetext[1]{...}
\end{table}

\begin{table}
    \caption{Column densities and column density ratios of water isotopologues.}\label{tab:N}
    \begin{tabular}{ccccccc} 
        \toprule
        Method & $N$(\ce{D2O}) & $N$(HDO)  & $N$(H$_2^{18}$O) & \ce{D2O}/\ce{H2O} & \ce{D2O}/HDO & $\frac{\mathrm{D_2O/HDO}}{\mathrm{HDO/H_2O}}$ \\
        & ($10^{13}$ cm$^{-2}$)  & ($10^{14}$ cm$^{-2}$) & ($10^{14}$ cm$^{-2}$) & ($\times 10^{-5}$) & $(\times 10^{-3})$ & \\
        \midrule
        Keplerian mask    & $4.2 \pm 1.2$ & $49.5 \pm 6.9$ & $23.7 \pm 5.7$ & $3.2 \pm 1.2$ & $8.5 \pm 2.8$ &  $2.3 \pm 1.0$ \\%& 
        Shifted spectrum  & $4.9 \pm 1.4$ & $56.4 \pm 7.7$ & $32.3 \pm 6.3$ & $2.7 \pm 1.0$ & $8.6 \pm 2.8$ &  $2.8 \pm 1.1$ \\%& ... \\
        \botrule
    \end{tabular}
    \footnotetext{All uncertainties include the absolute flux calibration uncertainty of ALMA, the uncertainty on the assumed excitation temperature, and the uncertainty on the \ce{^16O}/\ce{^18O} ratio.}
\end{table}

\FloatBarrier

\FloatBarrier

Correspondence and requests for materials should be addressed to Margot Leemker. 

% \clearpage
% \bibliographystylemain{unsrt}
% \bibliographymain{refs}

\renewcommand{\figurename}{Supplementary Data Figure}
\renewcommand{\tablename}{Supplementary Data Table}

\newpage
\FloatBarrier
\section*{Supplementary information}

\section{Observational details}

A full description of the self-calibration of the ALMA data can be found in the Methods section. The dates, number of antennas, configuration, and the bandpass, flux, and phase calibrators are summarized in Supplementary Table~\ref{tab:EBs}. In addition, the self-calibrated continuum emission of the V883~Ori disk is presented in Supplementary Figure~\ref{fig:cont}.

\begin{table}%[]
    \caption{ALMA observations covering the \ce{D2O} line.}\label{tab:EBs}
    \begin{tabular}{ccccccc} 
        \toprule
        Date & number of & min. baseline  & max. baseline & bandpass \& flux & phase  \\
             & antennas  & (m)           &  (km)       &  calibrator &  calibrator \\
        \midrule
        24 Dec. 2023 & 43 & 15.1 & 1.2 & J0423-0120 & J0607-0834 \\%& ...\footnotemark[1] \\
        28 Jul. 2024 & 45 & 15.1 & 1.4 & J0538-4405 & J0607-0834 \\%& ...\footnotemark[1] \\
        30 Jul. 2024 & 47 & 15.1 & 1.4 & J0538-4405 & J0607-0834 \\%& ...\footnotemark[1] \\
        \botrule
    \end{tabular}
    % \footnotetext{}
    %	\footnotetext[1]{...}
\end{table}

\begin{figure}[t]
    \centering
    \includegraphics[]{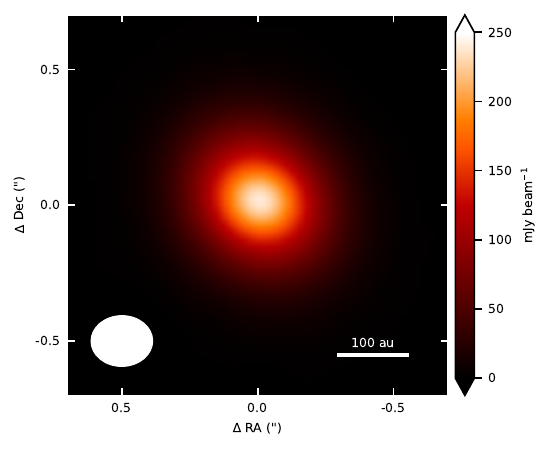} 
    \caption{Continuum image of the V883~Ori disk after self-calibration. The white ellipse in the bottom left corner indicates the beam of the observations. }  \label{fig:cont}
\end{figure}

\section{Water isotopologue ratios} \label{sec:D2OHDOHDOH2O}

The (\ce{D2O}/HDO) / (HDO/\ce{H2O}) has been suggested as an indicator to distinguish inheritance from reset in addition to the \ce{D2O}/\ce{H2O} ratio \citemain{Furuya2017}. 
The HDO/\ce{H2O} ratio is more accessible through observations than the (\ce{D2O}/HDO) / (HDO/\ce{H2O}) ratio as the former only requires observations of two water isotopologues. However, modeling of the ices in a two-dimensional collapsing core shows that the HDO/\ce{H2O} ratio in the case of reset can be similar to or even higher than the inherited value due to reformation of HDO ice from deuterium rich fragments on the ice and non-equilibrium gas-phase chemistry whose deuterated products freeze-out \citemain{Furuya2017}. 

To determine if water is inherited or reset, we follow a twofold argument. First of all, the water isotopologue ratios in the V883~Ori disk are compared to those in younger and older objects to look for similarity. Second, these ratios are compared to the model predictions by \citemain{Furuya2017}. Here, the most important uncertainties in the model predictions are discussed. 

\subsection{Modelling uncertainties}

The rotation of the envelope and number of chemically active monolayers in the chemical network typically only affect the water abundance by a factor of several \citemain{Furuya2017}. As the expected values for \ce{D2O}/\ce{H2O} and (\ce{D2O}/HDO) / (HDO/\ce{H2O}) between inheritance and reset generally differ by two orders of magnitude this is not expected to significantly affect the expected ratios.

The deuteration of water is only efficient at sufficiently low temperatures of $\lesssim 25$~K and densities above a few $10^4$~cm$^{-3}$ \citemain{Watson1976, Ceccarelli2014}. Therefore, the deuteration will be more (less) extreme than modelled if the temperature in the V883~Ori system were lower (higher) and the density was higher than in the models by \citemain{Furuya2017}. 

The abundance of HDO and \ce{D2O} in the case of inheritance can be further enhanced if the D/H in the gas increases at the time the water 
ice forms in the protostellar envelope or protoplanetary disk phase. This higher D/H in the gas only gets transferred to the water ice if the 
timescale for the gas-phase ion-neutral chemistry is faster than the timescale of the freeze-out of atomic oxygen. These timescales roughly match 
for a gas density of $10^4$~cm$^{-3}$ and a cosmic ray ionization rate of $10^{-17}$~s$^{-1}$ \citemain[e.g.,][]{Bergin2007}. Such low densities 
are unlikely at later evolutionary stages. For higher densities of $10^6$~cm$^{-3}$, the cosmic ray ionization rate needs to exceed 
$10^{-15}$~s$^{-1}$ to enhance the deuteration of water ice. As there are no indications for a cosmic ray ionization rate that is elevated by two 
orders of magnitude, an enhancement in the water deuteration in the protostellar envelope or protoplanetary disk phase of the V883~Ori system is 
not expected.

\begin{figure}[t]
    \centering
    \includegraphics{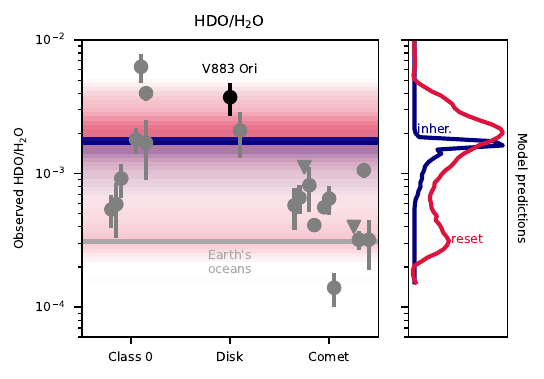} 
    \caption{ The HDO/\ce{H2O} ratio across different stages of star and planet formation. The measurement in the V883 Ori disk are presented in black and those in the Class~0 objects NGC~1333~IRAS~2A, NGC~1333~IRAS~4A-NW, NGC~1333~IRAS~4B, IRAS~16293-2422, B335, BHR71-IRS1, L483, the L~1551~IRS5 disk, and the comets in grey \protect\citemain{Jensen2019, Jensen2021b, Coutens2014, Tobin2023, Altwegg2017, Altwegg2015, Hagemann1970, Persson2014, Mandt2024, DeLaeter2003, BockeleeMorvan1998, Meier1998, Gibb2012, Villanueva2009, BockeleeMorvan2012, Biver2006, Biver2016, Lis2013, Hartogh2011, Lis2019, Andreu2023}. The errorbars represent the 1$\sigma$ uncertainty (s.d.) on the measured column density ratio in each source. The colored background and the histograms each normalized to the peak number of fluid parcels on the side indicate the expected water isotopologue ratios for inheritance where $\lesssim 10$\% of the \ce{H2O} ice is destroyed (blue) or reset where $\gtrsim70$\% of the \ce{H2O} is expected to be destroyed through photodissociation and photodesorption  in a model of a collapsing core (red; \protect\citemain{Furuya2017}). The red histogram is smoothed using a Savitzky-Golay filter with a window of 10 and an order of 3. }  \label{fig:inheritance_vs_reset_methods_HDOH2O}
\end{figure}

\subsection{HDO/\ce{H2O} ratio}
 Using the reimaged HDO and H$_2^{18}$O emission that was originally presented in \citemain{Tobin2023}, we derive the HDO/\ce{H2O} ratio to be $(3.7 \pm 1.0)\times 10^{-3}$. This is consistent within 1.2$\sigma$ with that of $(2.3\pm 0.6) \times 10^{-3}$ derived in \citemain{Tobin2023} and their conclusions.

A summary of the observed HDO/\ce{H2O} ratios across Class~0, the reanalyzed datapoint in the V883~Ori disk, and comets is presented in Supplementary Data Figure~\ref{fig:inheritance_vs_reset_methods_HDOH2O} \citemain{Jensen2019, Jensen2021b, Coutens2014, Tobin2023, Altwegg2017, Altwegg2015, Hagemann1970, Persson2014, Mandt2024, DeLaeter2003, BockeleeMorvan1998, Meier1998, Gibb2012, Villanueva2009, BockeleeMorvan2012, Biver2006, Biver2016, Lis2013, Hartogh2011, Lis2019}. %  refs 24, 26, 60, 61,  63-73 in Tobin2023 + the ones from my code
Recently it was found that the D/H ratio in water in the comet 67~P is lower than originally thought at a value of $(2.59\pm 0.36) \times 10^{-4}$ (corresponding to HDO/\ce{H2O} = $(5.18\pm 0.72) \times 10^{-4}$) due to the effect preferential adsorption of HDO on the dust grains \citemain{Mandt2024}. As the \ce{D2O} abundance has not been rederived taking the effect of the dust into account, we use previously derived HDO/\ce{H2O} ratio of $(1.06 \pm 0.14) \times 10^{-3}$ for a consistent analysis across all water isotopologue ratios measured in this comet \citemain{Altwegg2017}. Using the newly derived HDO/\ce{H2O} ratio in the comet 67~P would increase the (\ce{D2O}/HDO)/(HDO/\ce{H2O}) ratio by a factor of 4 assuming that the \ce{D2O} does not preferentially adsorb on the dust grains like HDO. This does not change our conclusions.
The color shading in the background of the left panel and the histogram on the right clearly demonstrate that the distributions of the expected HDO/\ce{H2O} ratio overlap for inheritance and reset and thus that this ratio does not distinguish inheritance from reset despite the similarity in the HDO/\ce{H2O} ratio across three evolutionary phases suggesting inheritance.

\begin{figure}[t]
    \centering
    \includegraphics{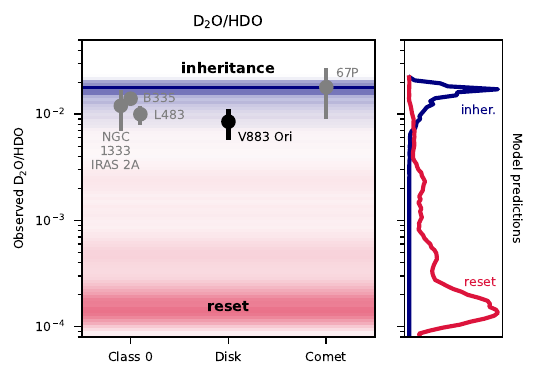} 
    \caption{The \ce{D2O}/HDO ratio across different stages of star and planet formation. The measurements in the V883 Ori disk are presented in black and those in the Class~0 objects NGC~1333~IRAS~2A, B335, L483, and the 67P comet in grey \protect\citemain{Coutens2014, Altwegg2015, Altwegg2017, Jensen2019, Jensen2021b}. The errorbars represent the 1$\sigma$ uncertainty (s.d.) on the measured column density ratio in each source. The colored background and the histograms each normalized to the peak number of fluid parcels on the side indicate the expected water isotopologue ratios for reset where $\gtrsim70$\% of the \ce{H2O} is expected to be destroyed through photodissociation and photodesorption  in a model of a collapsing core (red; \protect\citemain{Furuya2017}). The red histogram is smoothed using a Savitzky-Golay filter with a window of 10 and an order of 3.}\label{fig:inheritance_vs_reset_methods_D2OHDO}
\end{figure}

\subsection{\ce{D2O}/HDO ratio}
Another water isotopologue ratio is the \ce{D2O}/HDO ratio. This ratio in astronomical objects together with the model predictions by \citemain{Furuya2017} are presented in Supplementary Data Figure~\ref{fig:inheritance_vs_reset_methods_D2OHDO}. In case the material is inherited, the \ce{D2O}/HDO ratio is $\sim 2\times 10^{-2}$, whereas in case the material is reset, the \ce{D2O}/HDO ratio is $\sim 10^{-4}$, though high values up $\sim 1\times 10^{-2}$ are rare but still possible. The \ce{D2O}/HDO ratio in the V883~Ori disk is $(8.5 \pm 2.8) \times 10^{-3}$, much closer to the ratio expected for inheritance than for reset. In addition, this ratio is consistent within $1\sigma$ with that in the Class~0 sources L483 and NGC~1333~IRAS~2A, and with the comet 67~P and within $2\sigma$ of B335, the only other Class~0 source with an \ce{D2O}/HDO measurement. Therefore, the \ce{D2O}/HDO ratio in the V883~Ori disk is consistent with inheritance. 

\subsection{(\ce{D2O}/HDO) / (HDO/\ce{H2O}) ratio}
Similarly, the overlap between the (\ce{D2O}/HDO) / (HDO/\ce{H2O}) ratio for inheritance and reset is much less than that for the HDO/\ce{H2O} ratio and therefore this is a better tracer than the HDO/\ce{H2O} ratio \citemain{Furuya2017}. Model predictions show that the (\ce{D2O}/HDO) / (HDO/\ce{H2O}) ratio is $\sim 0.8-19$ if the material is inherited where the spread is due to e.g., variations in the initial ortho-to-para ratio of \ce{H2}, the abundances at the time at the time the prestellar core forms, and lifetime of the prestellar core before collapse \citemain{Furuya2016}. In the collapsing core models by \citemain{Furuya2017} the initial (\ce{D2O}/HDO) / (HDO/\ce{H2O}) ratio is 10. If material is reset, the expected value is $\sim 5 \times 10^{-2}$ and only few of the modeled stream lines predict a (\ce{D2O}/HDO) / (HDO/\ce{H2O}) above 1 in case of reprocessing (see Extended Data Figure 9) \citemain{Furuya2017}.

The observations of the (\ce{D2O}/HDO) / (HDO/\ce{H2O}) ratio in the V883~Ori disk together with those in younger Class~0 objects and the 67~P comet are presented in Extended Data Figure 9. All values are consistent with those for inheritance when variations in the initial conditions are considered \citemain{Furuya2016}. Only NGC~1333~IRAS~2A and 67~P are consistent with the model prediction for inheritance by \citemain{Furuya2017} within $1\sigma$. The two remaining Class~0 sources and the V883~Ori disk have a somewhat lower value of $\sim2-3$, possibly due to e.g., a different ortho-to-para ratio of \ce{H2} and thus a different reservoir of chemical energy in the initial phases or due a different evolution of the early phases such as the time before the onset of collapse of the prestellar core and the visual extinction of the cloud when the initial conditions of the chemical network are initialized \citemain{Furuya2016}. The similarity in the observed (\ce{D2O}/HDO) / (HDO/\ce{H2O}) ratio across Class~0 objects and the V883~Ori disk, together with the spread in possible values for inheritance strongly suggests that ice is preserved from the earliest phases of star and planet formation.

\section{Systematic uncertainties} \label{sec:caveats}
In addition to the effect of the continuum optical depth potentially leading to an underestimation of the water isotopologue ratios discussed in the main text, other systematic effects could affect the derived ratios. %These are discussed in the following subsections. 

\subsection{Excitation temperature} \label{sec:Tex}
As only a single line of each water isotopologue is analyzed in this work, their measured column densities are derived using an assumed excitation temperature of $199\pm 42$~K derived from a rotational diagram analysis of two HDO lines by \citemain{Tobin2023}. As one HDO line at 241~GHz used in that analysis is severely blended with two COMs, the derived excitation temperature is somewhat uncertain. The temperature derived from these two HDO lines is higher than the typically assumed 120~K temperature for COMs in this disk based on their desorption temperature of $\sim 100$~K \citemain{Lee2019, Jeong2024, Yamato2024, vantHoff2018} and lower than the excitation temperature of $343 \pm 102$~K for \ce{CH3CN}, the warmest COM seen in this disk \citemain{Jeong2024}.  % see their table 4
Supplementary Data Table~\ref{tab:Tex} summarizes the effect of the assumed excitation temperature on the derived \ce{D2O}/\ce{H2O} and the (\ce{D2O}/HDO) / (HDO/\ce{H2O}) ratios by varying the temperature by $\pm 100$~K, the approximate range of temperatures found for COMs in this disk. These ratios increase with temperature, driving the results further into inheritance if the gas is warmer than assumed. Even if the gas were only 99~K, the \ce{D2O}/\ce{H2O} ratio would be fully consistent with inheritance of material. In addition, the (\ce{D2O}/HDO) / (HDO/\ce{H2O}) ratio at this temperature is consistent within $1\sigma$ with those in Class~0 objects tracing pristine material.

\begin{table}%[]
    \caption{Effect of the excitation temperature on the \ce{D2O}/\ce{H2O} and the (\ce{D2O}/HDO) / (HDO/\ce{H2O}) ratios.}\label{tab:Tex}
    \begin{tabular}{ccc} 
        \toprule
        $T_{\rm rot}$ (K) & \ce{D2O}/\ce{H2O} & (\ce{D2O}/HDO) / (HDO/\ce{H2O}) \\
        \midrule
         99 & $(1.3 \pm 0.8)\times 10^{-5}$ & $1.4 \pm 0.9$ \\%& ...\footnotemark[1] \\
        199 & $(3.2\pm 1.2)\times 10^{-5}$ & $2.3 \pm 1.0$ \\%& ...\footnotemark[1] \\
        299 & $(4.3\pm  1.4)\times 10^{-5}$ & $2.7 \pm 1.0$ \\%& ...\footnotemark[1] \\			
        \botrule
    \end{tabular}
    \footnotetext{All uncertainties include the absolute flux calibration uncertainty of ALMA, the uncertainty on the assumed excitation temperature, and the uncertainty on the \ce{^16O}/\ce{^18O} ratio.}
    %	\footnotetext[1]{...}
\end{table}

\subsection{Optical depth} \label{sec:tau}

The column densities of the water isotopologues are derived in the assumption of optically thin emission. The optical depth at the line center $\tau$ can be approximated as \citemain{Goldsmith1999, Loomis2018}:
\begin{align}
    \tau = \frac{A_{ul} c^3}{8\pi \nu^3 \Delta V}N_u \left (e^{h\nu / kT_{\rm rot}} -1 \right ) 
\end{align}
with $\nu$ the frequency of the line and $\Delta V$ the thermal linewidth. This expression approximates the peak of the normalized line profile as $1/\Delta V$ which is very close to that for a Gaussian line profile $2\sqrt{\ln{2}}/(\sqrt{\pi} \Delta V)$. The thermal linewidth is defined as:
\begin{equation}
    % \Delta f = \sqrt{\frac{8kT_{\rm rot} \ln 2 }{m c^2}}f_0
    \Delta V = \sqrt{\frac{8kT_{\rm rot} \ln 2 }{m}}      
\end{equation}
with $m$ the mass of the molecule. For all three water isotopologues, this results in an FWHM linewidth of $\sim 0.7$~km~s$^{-1}$ at a temperature of $199$~K. This is slightly narrower than the line width in the shifted spectra because the finite resolution of the data and possible differences in the true emitting geometry set by the stellar mass, inclination, position angle, and assumed emitting height. An overview of the optical depth estimates is presented in Supplementary Data Table~\ref{tab:tau}. 

For the fiducial temperature of 199~K, the thermal linewidth and derived column densities of the upper energy levels, all lines analyzed in this work are optically thin. The HDO line has the highest optical depth at 0.14, which results in a 7\% correction factor on the column density using the optical depth correction factor $\tau / (1-e^{-\tau})$ \citemain{Goldsmith1999, Loomis2018}. At a lower temperature of 99~K, the HDO line has an optical depth of 0.54 which leads to a correction factor of 29\% on the HDO column density. This lowers the (\ce{D2O}/HDO) / (HDO/\ce{H2O}) ratio to 1.5 %(0.042/(1-e^(-1*0.042)) * 0.08/(1-e^(-1*0.08)) )  / (0.54/(1-e^(-1*0.54)) )^2 = 0.6344
which is at the boundary of 1 between inheritance and reset. At a higher temperature of 299~K, all lines are fully optically thin. The best tracer to distinguish inheritance from reset used in this work, the \ce{D2O}/\ce{H2O} ratio, is not affected by the optical depth of HDO. Therefore, the observed \ce{D2O} and H$_2^{18}$O emission remain consistent with the inheritance scenario even if the temperature is different from $199 \pm 42$~K. 

Finally, the effect of the assumed emitting region for the emission is summarized in the final column of Supplementary Data Table~\ref{tab:tau}. The fiducial region of $0\farcs4$ radius in the frame of the disk is motivated by the HDO and H$_2^{18}$O emission that is seen out to that distance \citemain{Tobin2023}. As the midplane water snowline is located at $\sim$80~au ($0\farcs2$) indicated by the steep drop off in the HDO and H$_2^{18}$O column densities, the smallest physically motivated emitting region is an elliptical annulus between $0\farcs1$ and $0\farcs2$. The inner boundary is set by the optically thick dust that hides the line emission out to 40~au ($0\farcs1$). 

The area of the annulus is a factor of 5.3 smaller than the fiducial region, increasing the derived column densities by that same factor. As all column densities change by that factor under the assumption of optically thin emission, the derived line ratios remain unaffected. The optical depths of the water isotopologue lines on the other hand increases by a factor of 5.3. For this small emitting region, the HDO emission becomes optically thick with an optical depth of $9.8\times 10^{-1}$, making the water isotopologue ratios involving HDO hard to interpret as they are very sensitive to the precise optical depth and because the HDO possibly emits from a higher layer in the disk than the \ce{D2O} and H$_2^{18}$O. 
The \ce{D2O} and H$_2^{18}$O on the other hand remain approximately optically thin with optical depths of $7.6\times 10^{-2}$ and $1.5\times 10^{-1}$, respectively. Therefore, even in this case the \ce{D2O}/\ce{H2O} ratio can be used to infer that water is inherited.

\begin{table}%[]
    \caption{Optical depth of the lines analyzed in this work.}\label{tab:tau}
    \begin{tabular}{ccccc} 
        \toprule
        molecule & $\tau$ & $\tau$ & $\tau$ & $\tau$ \\
         & $T_{\rm rot}= 199$ K (fid.) & $T_{\rm rot}= 99$ K & $T_{\rm rot}= 299$ K & $T_{\rm rot}= 199$ K \& \\        %small emitting region
         & & & & small emitting region \\
        \midrule
         \ce{D2O}    & $1.4\times 10^{-2}$ & $4.2\times 10^{-2}$ & $7.6\times 10^{-3}$ & $7.6\times 10^{-2}$ \\%& ...\footnotemark[1] \\
         HDO         & $1.8\times 10^{-1}$ & $5.4\times 10^{-1}$ & $9.9\times 10^{-2}$ & $9.8\times 10^{-1}$\\%& ...\footnotemark[1] \\
         H$_2^{18}$O & $2.8\times 10^{-2}$ & $8.0\times 10^{-2}$ & $1.5\times 10^{-2}$ & $1.5\times 10^{-1}$ \\%& ...\footnotemark[1] \\  
        \botrule
    \end{tabular}
    %	\footnotetext[1]{...}
\end{table}

\subsection{Ortho-to-para ratios}
The \ce{D2O} and H$_2^{18}$O molecules both have ortho and para states due to the presence of two deuterium and two hydrogen atoms, respectively. Observations of \ce{D2O} in the IRAS~16293-2422 cold protostellar envelope show that the observed value is consistent with the statistical value of 2 \citemain{Vastel2010}. In addition, observations of the main water isotopologue show a similar trend with most observations being consistent with the statistical value of 3 \citemain[and references therein]{Faure2019}. If the true ortho-to-para ratio (OPR) is different than the assumed statistical then the column density derived is scaled by a factor of $\rm (1+OPR)/3$ and $\rm (1+OPR)/4$, respectively. If the OPR is at a low value of 1 for both molecules, then the derived \ce{D2O}/\ce{H2O} increases to $(4.2\pm 1.6)\times 10^{-5}$ and (\ce{D2O}/HDO) / (HDO/\ce{H2O}) ratio decreases to $0.8 \pm 0.3$. Even in this case, the \ce{D2O}/\ce{H2O} ratio remains consistent with inheritance.

\subsection{Masing} \label{sec:masing}

The column densities of the water isotopologues are derived under the assumption of LTE. However, a recent analysis by \citemain{Faure2024} showed that the HDO 225~GHz and the H$_2^{18}$O lines are possibly weakly masing in a large region in the V883~Ori disk. At the moment, detailed calculations of the masing conditions of the \ce{D2O} line are not reported. Still, the critical density for transitions of a non-linear, polyatomic molecule like the \ce{D2O} $1_{1,0}-1_{0,1}$ transition at 200~K is $9.9 \times 10^{5}$~cm$^{-3}$ using the collisional rate coefficients by \citemain{Faure2024}. This is much lower than the expected densities in the V883 Ori disk. Therefore, the \ce{D2O} $1_{1,0}-1_{0,1}$ line is expected to be in LTE.

If the HDO and H$_2^{18}$O lines are indeed weakly masing as suggested by \citemain{Faure2024}, their column densities are overestimated due to the assumption of LTE for the rotational diagram analysis. In addition, the relation between column density and observed flux becomes highly sensitive to the local conditions. Therefore, the (\ce{D2O}/HDO) / (HDO/\ce{H2O}) is not a meaningful measure in this case. However, the \ce{D2O}/\ce{H2O} ratio can still be used as the potential masing of the H$_2^{18}$O line only drives this ratio further into the inheritance regime whereas the \ce{D2O} line is expected to be in LTE.

\bibliographystylemain{unsrt}
\bibliographymain{refs}% common bib file

\newpage
\FloatBarrier

\section{Extended Data}

\begin{figure}[t]
    \centering
    \includegraphics[width=1\linewidth]{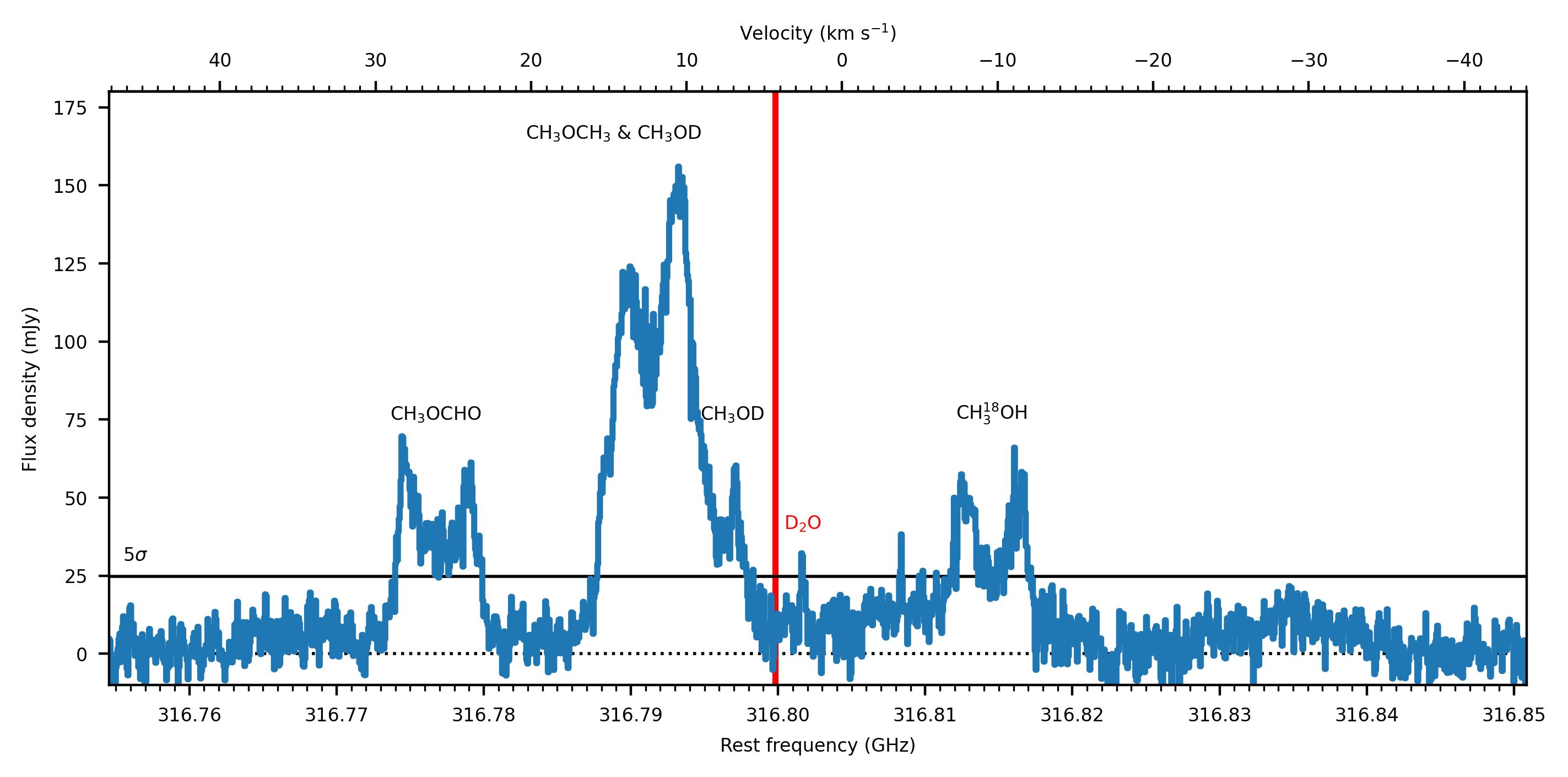}
    \caption{Integrated spectrum of \ce{D2O} in the V883~Ori disk extracted from a $0\farcs4 \times 0\farcs31$ (semi-axes) region. The red vertical line indicates the center of the double peaked \ce{D2O} emission due to the Keplerian rotation of the disk. Other bright emission lines  are indicated in black. The solid horizontal line indicates the 5$\sigma$ noise level measured in off-source regions in the image cube.} \label{fig:spec_not_stacked}
\end{figure}

\begin{figure*}[t]
    \centering
    \includegraphics[width=1\linewidth]{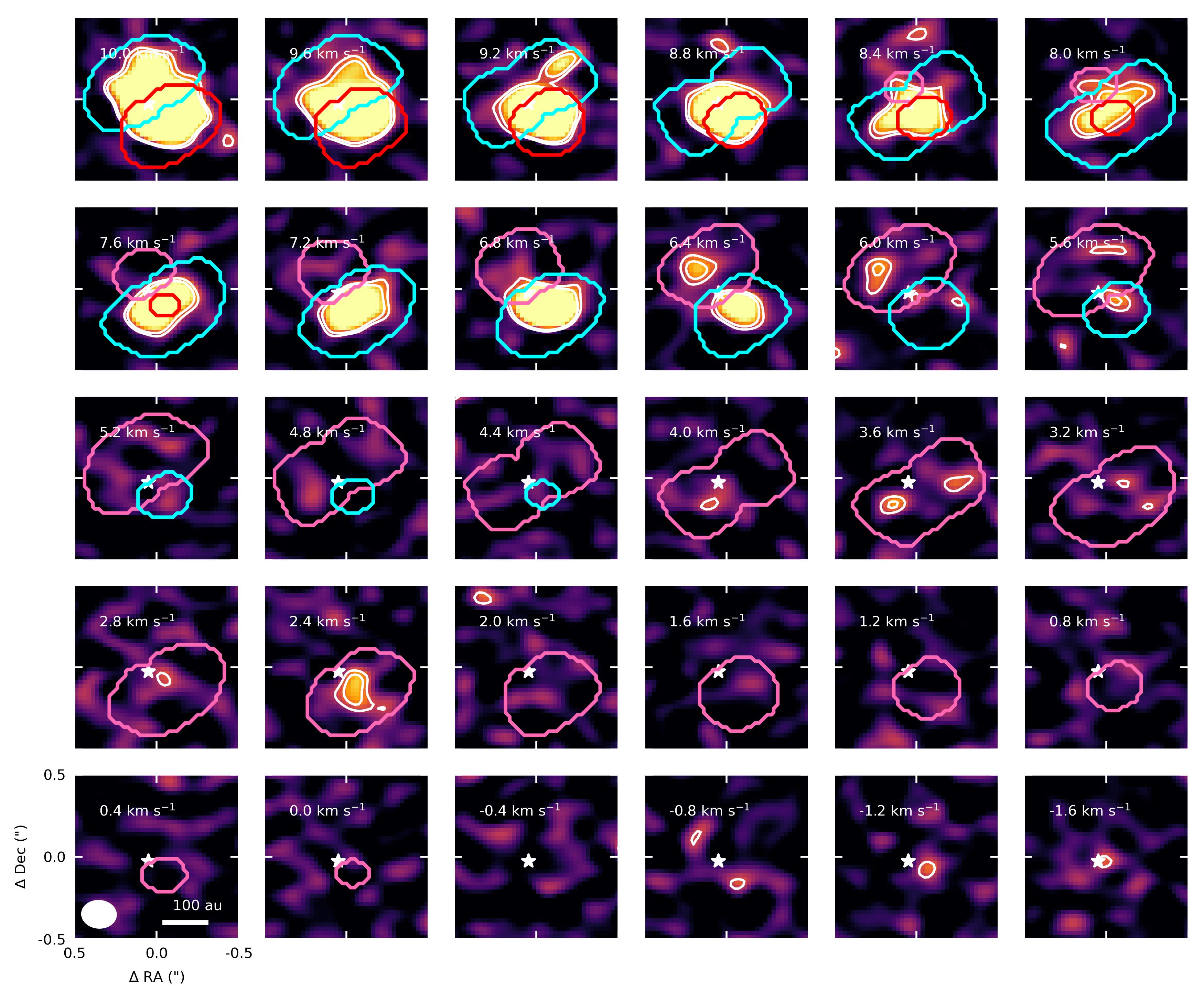}
    \caption{Channel maps of the \ce{D2O} emission. The pink contour indicates the region where \ce{D2O} is expected to emit if the emission is Keplerian. The white contours indicate the 3 and 4$\sigma$ level for the emission of \ce{D2O} and the neighbouring COMs. The cyan and red contours indicate the Keplerian mask where the neighboring lines that are blended with \ce{D2O} are expected to emit.} \label{fig:D2O_channel_maps_kep_masks}
\end{figure*}

\begin{figure*}[t]
    \centering
    \includegraphics[width=1\linewidth]{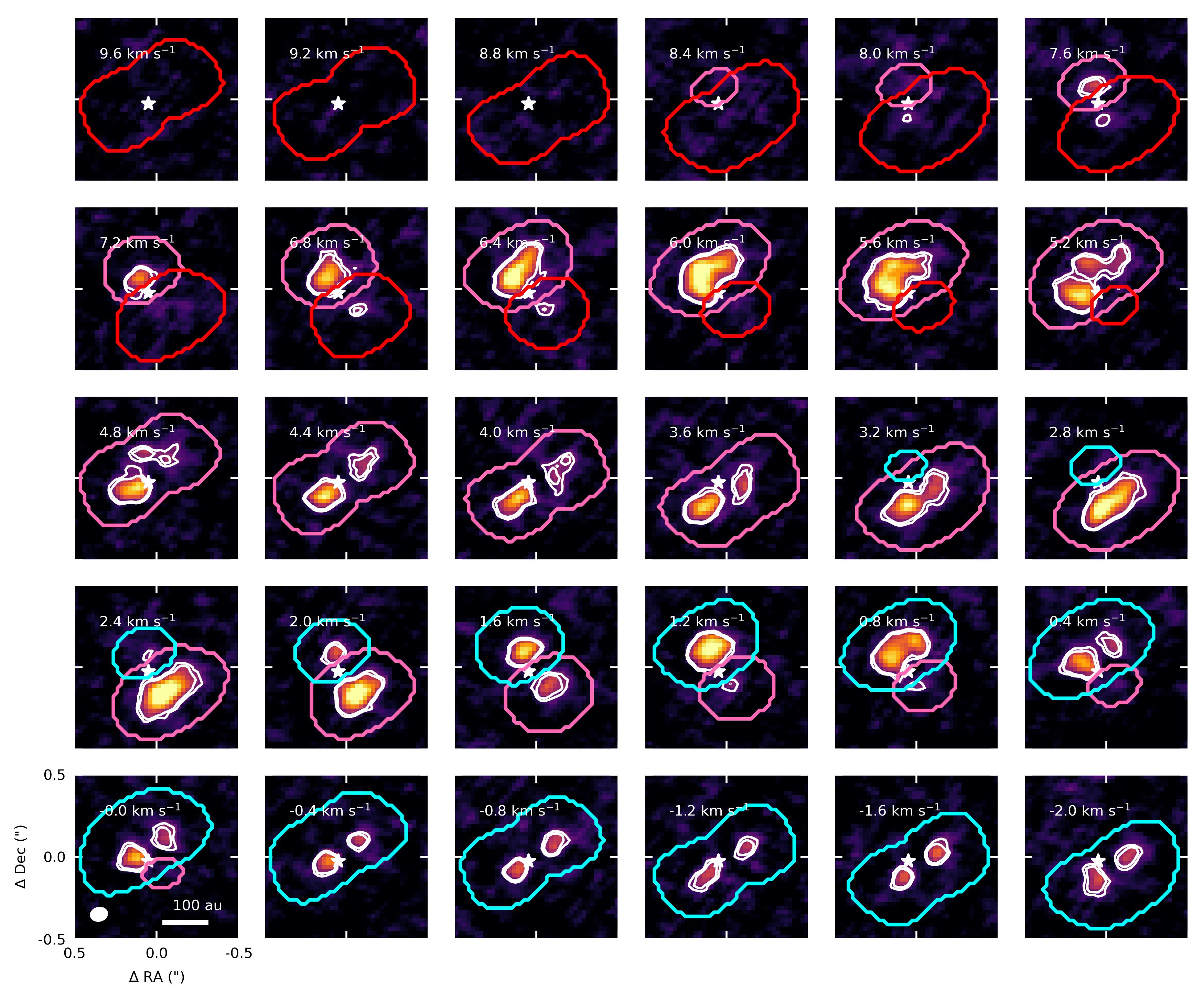}
    \caption{Channel maps of the HDO emission. The pink contour indicates the region where HDO is expected to emit if the emission is Keplerian and the white contours indicate the 3 and 4$\sigma$ level in these channels. The cyan and red contours indicate the Keplerian mask where the neighboring lines that are blended with HDO are expected to emit.} \label{fig:HDO_channel_maps}
\end{figure*}		

\begin{figure*}[t]
    \centering
    \includegraphics[width=1\linewidth]{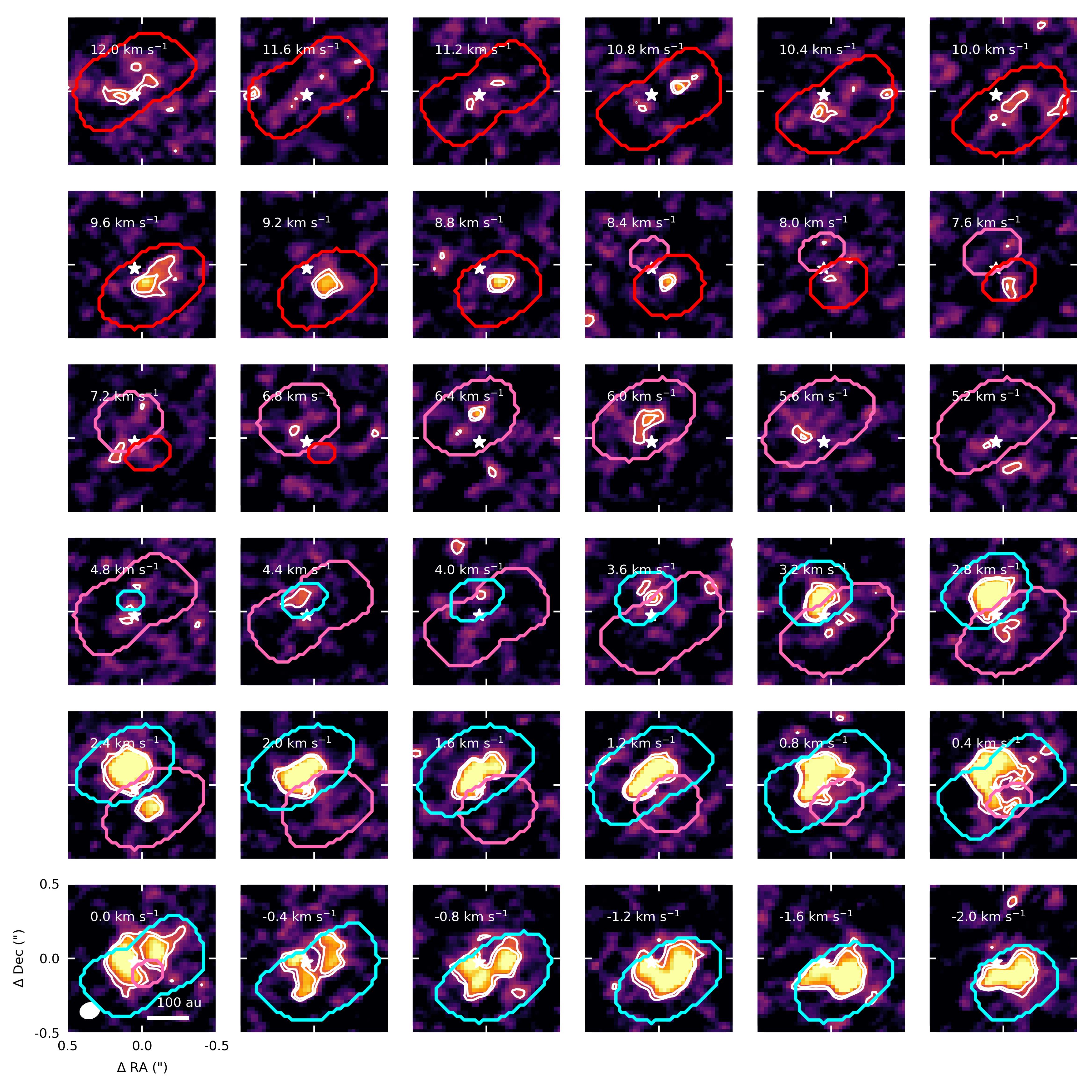}
    \caption{Channel maps of the H$_2^{18}$O emission. The pink contour indicates the region where H$_2^{18}$O is expected to emit if the emission is Keplerian and the white contours indicate the 3 and 4$\sigma$ level in these channels. The cyan and red contours indicate the Keplerian mask where the neighboring lines that are blended with H$_2^{18}$O are expected to emit.} \label{fig:H218O_channel_maps}
\end{figure*}

\begin{figure}[t]
    \centering
    \includegraphics[width=1\linewidth]{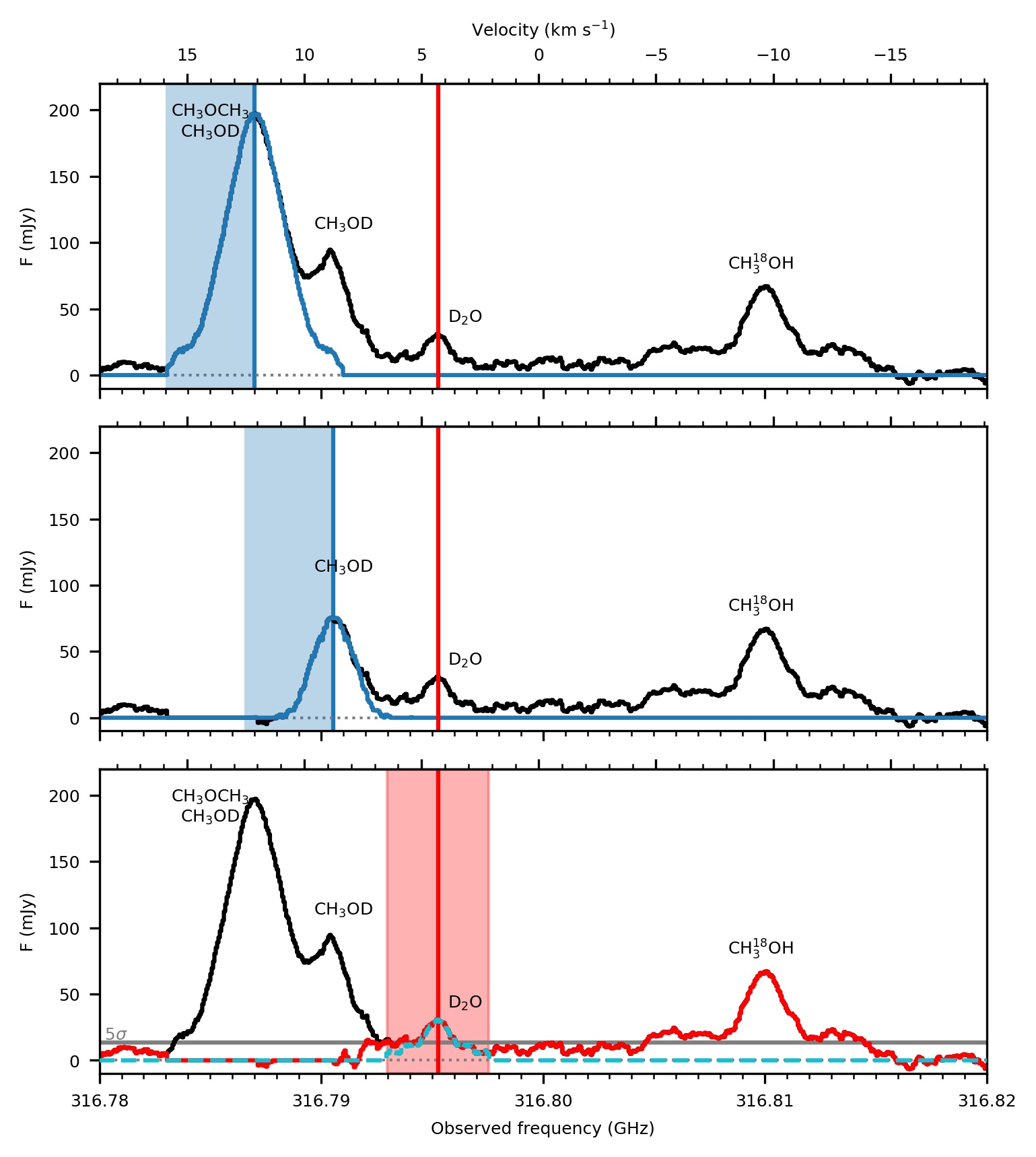}
    \caption{\ce{D2O} flux measurement using the shifted spectra. The lines blended with the \ce{D2O} line are subtracted by extracting the line profile of the clean, low frequency side of the blend (blue shaded region), mirroring this in the frequency of the line and subtracting. The top panel shows the model for the \ce{CH3OD} and \ce{CH3OCH3} in blue. In the middle panel, this model is subtracted from the data (black) and the new model for the \ce{CH3OD} is shown in blue. The \ce{D2O} line is indicated with the red vertical line in the bottom panel. The cyan line profile in this panel is used to measure the \ce{D2O} flux. }\label{fig:flux_measurement_D2O_stacked}
\end{figure}

\begin{figure}[t]
    \centering
    \includegraphics[width=1\linewidth]{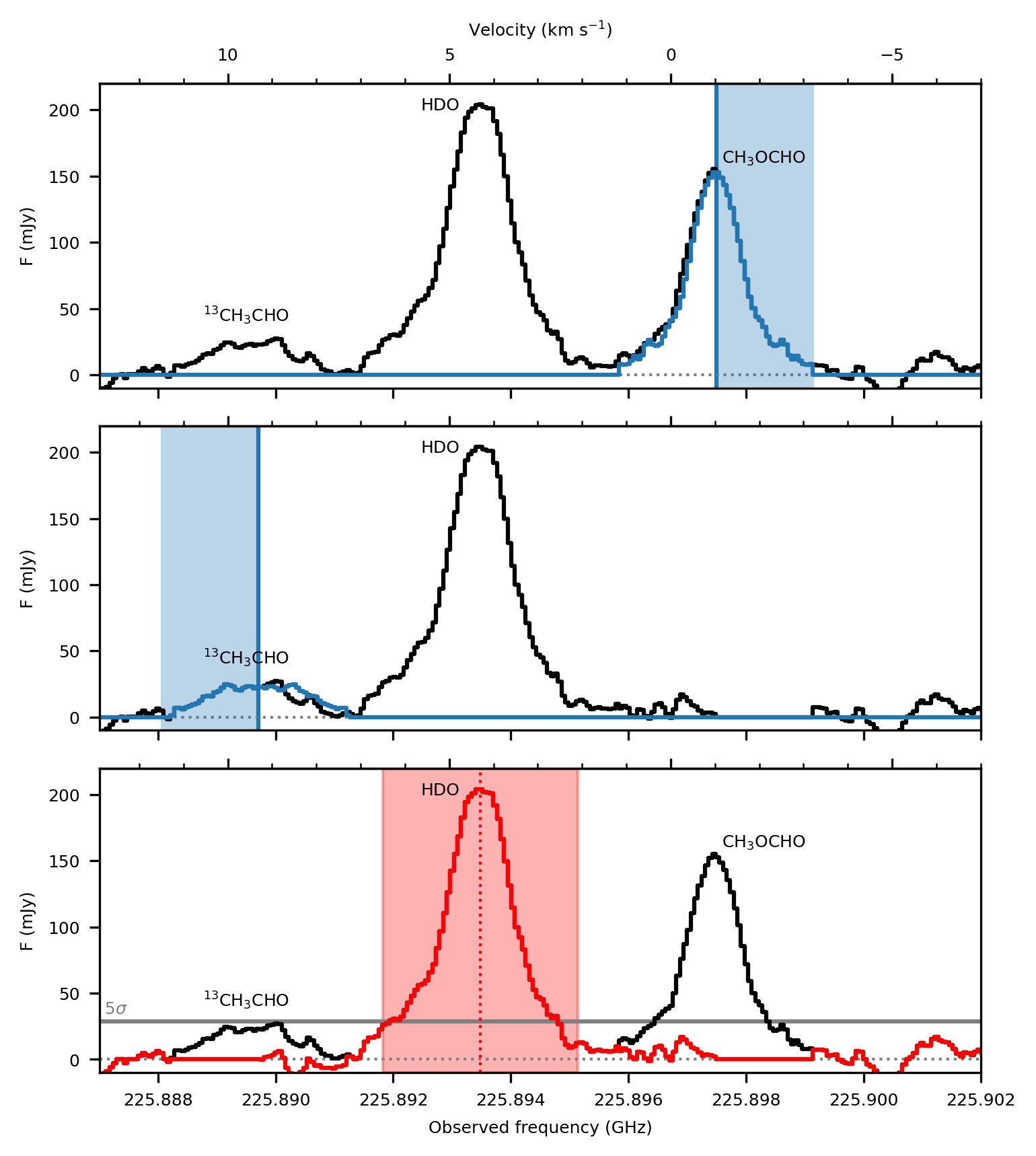}
    \caption{HDO flux measurement using the shifted spectra. The lines blended with the HDO line are subtracted by extracting the line profile of the clean, low frequency side of the blend (blue shaded region), mirroring this in the frequency of the line and subtracting. The top panel shows the model for the \ce{CH3OCHO} in blue. In the middle panel, this model is subtracted from the data (black) and the new model for the \ce{^13CH3CHO} is shown in blue. The HDO line is indicated with the red vertical line in the bottom panel.} \label{fig:flux_measurement_HDO_stacked}
\end{figure}

\begin{figure}[t]
    \centering
    \includegraphics[width=0.8\linewidth]{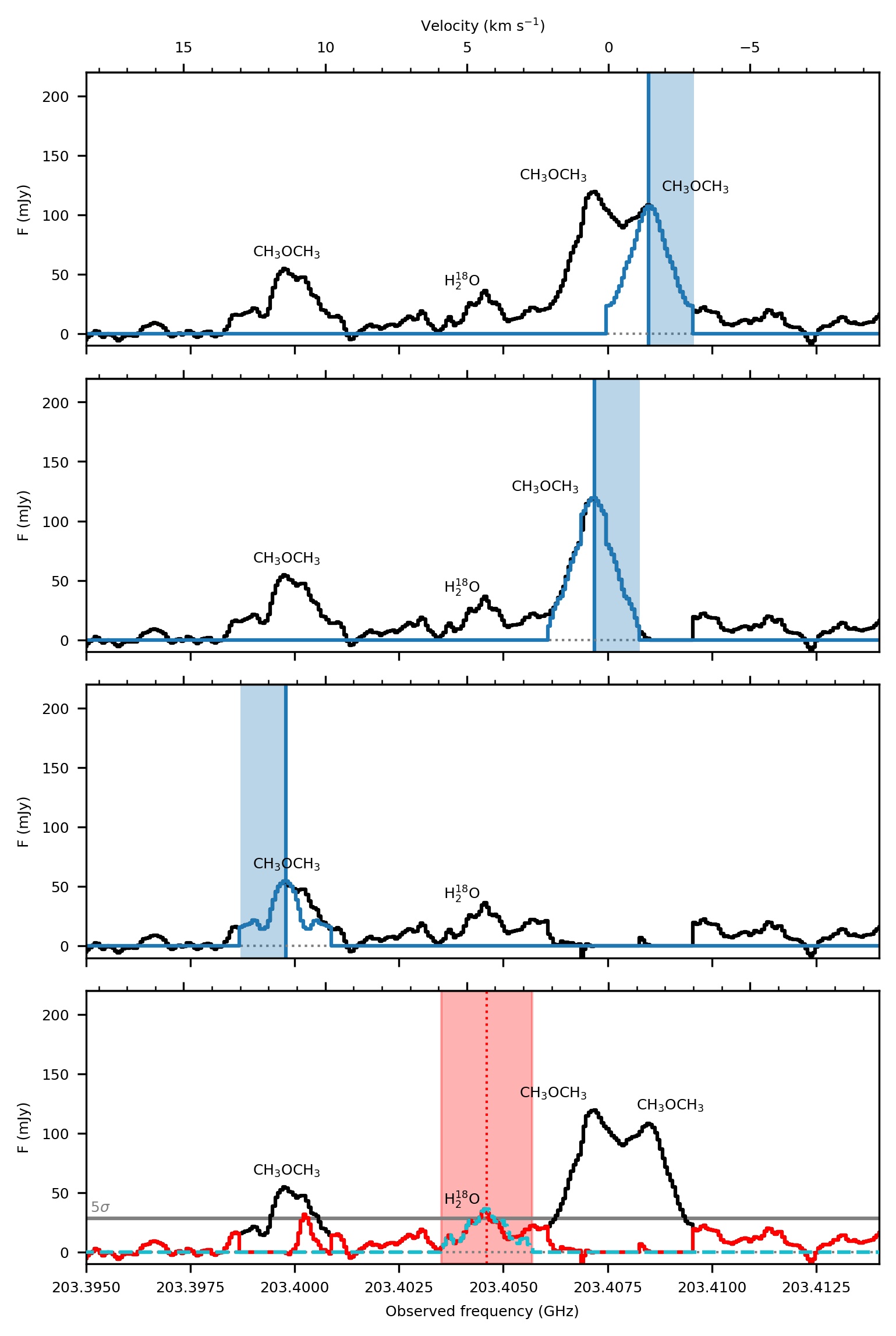}
    \caption{H$_2^{18}$O flux measurement using the shifted spectra. The lines blended with the H$_2^{18}$O line are subtracted by extracting the line profile of the clean side of the blend (blue shaded region), mirroring this in the frequency of the line and subtracting. The top three panels demonstrate this process for three \ce{CH3OCH3} lines. The cyan line profile in the bottom panel is used to measure the H$_2^{18}$O flux. } \label{fig:flux_measurement_H218O_stacked}
\end{figure}

\begin{figure}[t]
    \centering
    \includegraphics[width=0.97\linewidth]{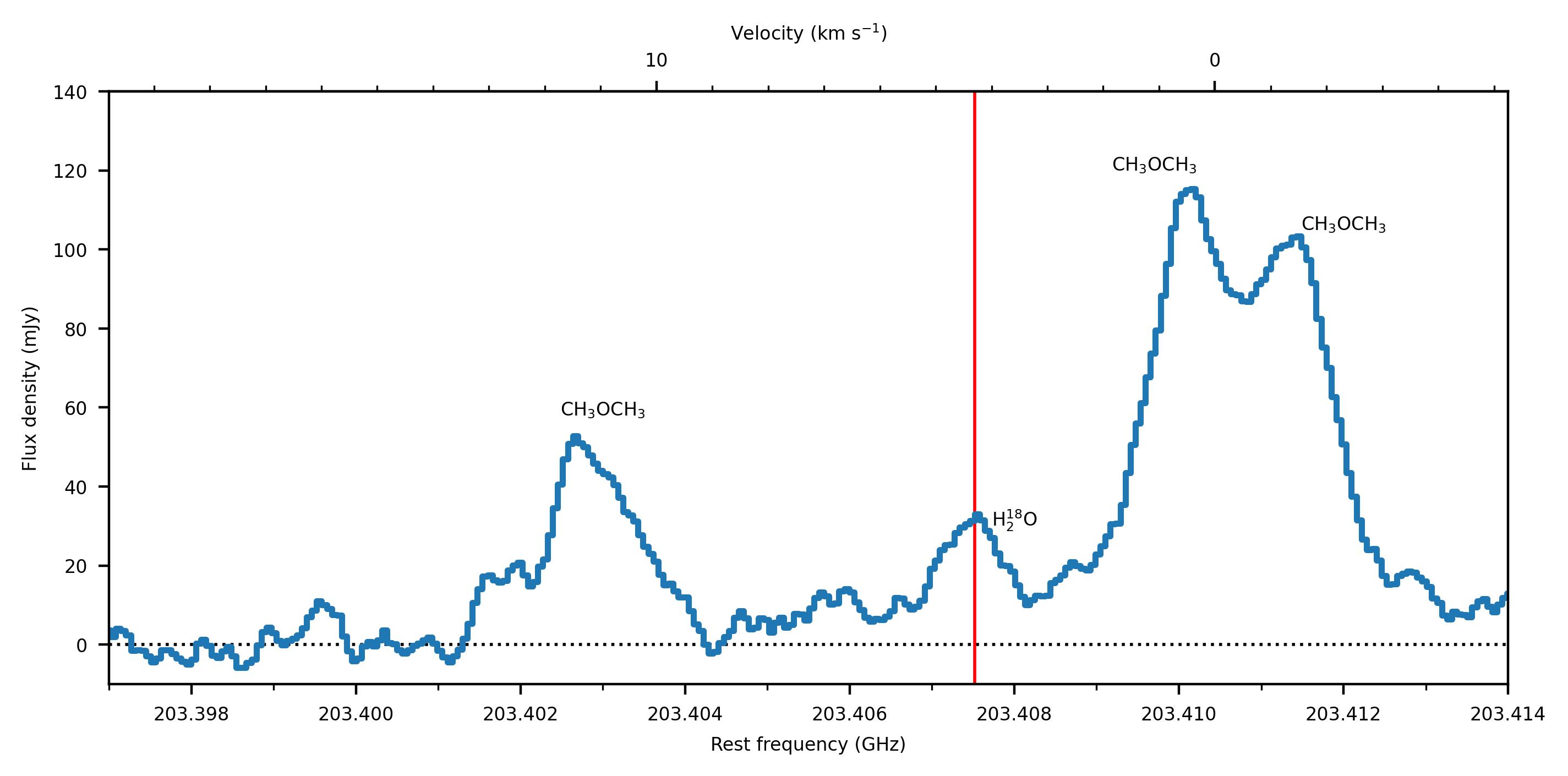}
    \caption{Shifted spectrum of the H$_2^{18}$O line extracted from an elliptical region with a semi-major axis of $0\farcs35$. This spectrum clearly shows that the H$_2^{18}$O line is detected in this disk.} \label{fig:flux_measurement_H218O_stacked_small}
\end{figure}

\begin{figure}[t]
    \centering
    \includegraphics{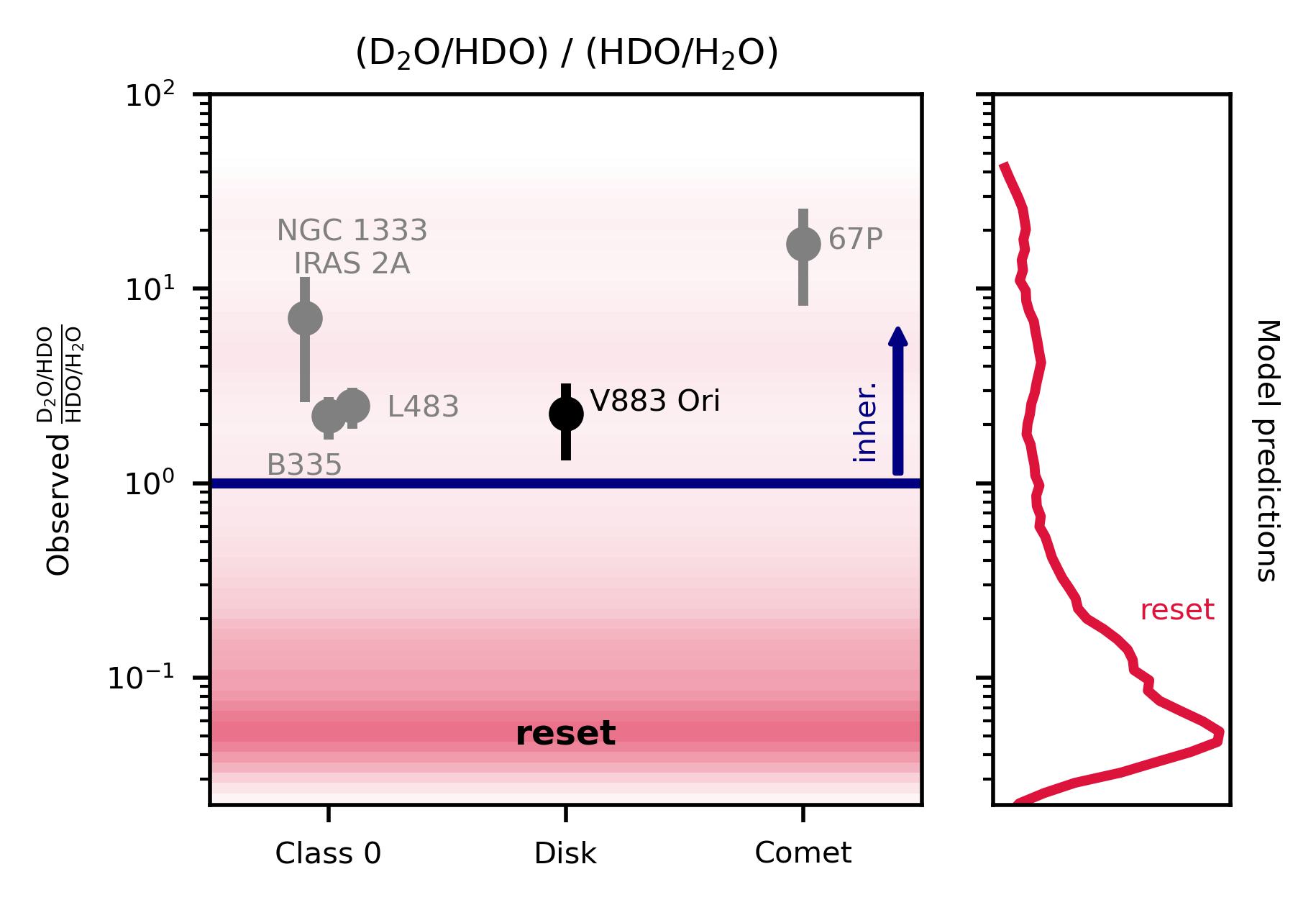} 
    \caption{ The (\ce{D2O}/HDO)/(HDO/\ce{H2O}) ratio across different stages of star and planet formation. The measurements in the V883 Ori disk are presented in black and those in the Class~0 objects NGC~1333~IRAS~2A, B335, L483, and the 67P comet in grey \protect\citemain{Coutens2014, Altwegg2015, Altwegg2017, Jensen2019, Jensen2021b}. The errorbars represent the 1$\sigma$ uncertainty (s.d.) on the measured column density ratio in each source. The colored background and the histograms each normalized to the peak number of fluid parcels on the side indicate the expected water isotopologue ratios for reset where $\gtrsim70$\% of the \ce{H2O} is expected to be destroyed through photodissociation and photodesorption  in a model of a collapsing core (red; \protect\citemain{Furuya2017}). The red histogram is smoothed using a Savitzky-Golay filter with a window of 10 and an order of 3. The horizontal blue line and arrow indicate an approximate boundary between inheritance ((\ce{D2O}/HDO)/(HDO/\ce{H2O}) $\gtrsim1$) and reset ((\ce{D2O}/HDO)/(HDO/\ce{H2O}) $\lesssim 1$) \protect\citemain{Furuya2017}.}  \label{fig:inheritance_vs_reset_methods}
\end{figure}

\end{document}